\documentclass[iop,apj,tighten,twocolappendix,numberedappendix]{emulateapj}
\usepackage{apjfonts}

\usepackage{graphicx}
\usepackage{amsmath}
\usepackage{amssymb}
\usepackage{color}
\usepackage{mathtools}
\usepackage[breaklinks,colorlinks,citecolor=blue,linkcolor=red]{hyperref} 
\usepackage[all]{hypcap}

\newcommand\msun{\, \rm M_\odot}

\begin{document}

\title{Tidal Disruption Events and Gravitational Waves from Intermediate-mass Black Holes in Evolving Globular Clusters Across Space and Time}

\author{Giacomo Fragione\altaffilmark{1}, Nathan W. C. Leigh\altaffilmark{2,3,4}, Idan Ginsburg\altaffilmark{5}, and Bence Kocsis\altaffilmark{6}}
 \affil{$^1$Racah Institute for Physics, The Hebrew University, Jerusalem 91904, Israel} 
 \affil{$^2$Department of Astrophysics, American Museum of Natural History, New York, NY 10024, USA}
\affil{$^3$Department of Physics and Astronomy, Stony Brook University, Stony Brook, NY 11794-3800, USA}
\affil{$^4$Center for Computational Astrophysics, Flatiron Institute, 162 Fifth Avenue, New York, NY 10010, USA}
  \affil{$^5$Astronomy Department, Harvard University, 60 Garden St, Cambridge, MA 02138, USA}
  \affil{$^6$Institute of Physics, E\"{o}tv\"{o}s University, P\'azm\'{a}ny P. s. 1/A, Budapest, 1117, Hungary}


\begin{abstract}
We present a semi-analytic model for self-consistently evolving a population of globular clusters (GCs) in a given host galaxy across cosmic time. We compute the fraction of GCs still hosting intermediate-mass black holes (IMBHs) at a given redshift in early and late type galaxies of different masses and sizes, and the corresponding rate of tidal disruption events (TDEs), both main-sequence (MS) and white dwarf (WD) stars. We find that the integrated TDE rate for the entire GC population can exceed the corresponding rate in a given galactic nucleus and that $\sim 90$\% of the TDEs reside in GCs within a maximum radius of $\sim 2-15$ kpc from the host galaxy's center. This suggests that observational efforts designed to identify TDEs should not confine themselves to galactic nuclei alone, but should also consider the outer galactic halo where massive old GCs hosting IMBHs would reside. Indeed, such off-centre TDEs as predicted here may already have been observed. MS TDE rates are more common than WD TDE rates by a factor 30 (100) at $z\leq 0.5$ ($z=2$). We also calculate the rate of IMBH-SBH mergers across cosmic time, finding that the typical IMRI rate at low redshift is of the order of $\sim 0.5-3$ Gpc$^{-3}$ yr$^{-1}$, which becomes as high as $\sim 100$ Gpc$^{-3}$ yr$^{-1}$ near the peak of GC formation. Advanced LIGO combined with VIRGO, KAGRA, ET and \textit{LISA} will be able to observe the bottom-end and top-end of the IMBH population, respectively.
\end{abstract}


\section{Introduction}

While the existence of supermassive (SMBHs, $M\gtrsim 10^5\ \mathrm{M}_{\odot}$) and stellar-mass black holes (SBHs, $10\ \mathrm{M}_{\odot}\lesssim M\lesssim 100\ \mathrm{M}_{\odot}$) has been confirmed, there is only circumstantial observational evidence for the presence of intermediate-mass black holes (IMBHs) ($100\ \mathrm{M}_{\odot}\lesssim M\lesssim 10^5\ \mathrm{M}_{\odot}$). One place to look for IMBHs is at the centres of globular clusters (GCs), assuming that the observed $M_{SMBH}-\sigma$ relation ($\sigma$ is the velocity dispersion) holds also for the range of IMBH masses \citep[e.g.][]{kruijssen13,mer13}. 

GCs are one of the most promising environments to form an IMBH. Several studies showed that the most massive stars may segregate and merge in the core of the cluster, forming a massive growing object of similar mass to an IMBH \citep{mil02b,por02,fre06,gie15}. Yet, the origin and formation of IMBHs is still a highly debated topic, and other mechanisms have been proposed such as the direct collapse of Pop III stars \citep{mad01} and fragmentation in disks surrounding an SMBH \citep{McKernan+2012,McKernan+2014}. If IMBHs reside at the centres of GCs, they interact with the host cluster environment and influence its evolution and composition \citep{lei14,bau17,frgual18}. No strong dynamical evidence has been found for the existence of IMBHs in GCs to date, in part due to a lack of sufficiently high-resolution data for the innermost GC regions. In spite of this, the hunt for IMBHs is still very active \citep{cann18,chili18,trem18,wrobel18}. Recently, \citet{lin18} observed a TDE event consistent with an IMBH in an off-centre star cluster, at a distance of $\sim 12.5$ kpc from the center of the host galaxy. \citet{chenshen18} studied the long-term accretion and the observational consequence of such TDE, and found a nice agreement between their accretion model and the observed TDE, thus supporting the existence of the IMBH engine. Based on N-body modelling, two clusters have been claimed to host an IMBH in the Milky Way, i.e. $\omega$ Cen \citep{bau17} and $47$ Tuc \citep{kiz17}.

The Milky Way galactic centre may host several IMBHs in its nuclear star cluster, possibly delivered by inspiraling clusters \citep{por02,ebi01,mas14,agu18,fragk18}, whose dynamical effects and/or nHz-frequency gravitational waves may be detected in the future \citep{gua09,gua10,koc12,lut13,mer13,lei14,mac16,fragk18}.

Gravitational wave (GW) astronomy will help in the hunt for the first IMBHs to be discovered \textit{and} confirmed. IMBH-SBH binaries may form in the core of GCs and may merge as IMRIs (intermediate mass ratio inspirals), which represent a down-sized version of extreme mass-ratio inspirals, the inspiral of a stellar BH into an SMBH \citep{hop06,ama07,man08,kon13,frlei18}. Present and upcoming facilities, such as LIGO\footnote{\url{http://www.ligo.org}}, the Einstein Telescope\footnote{\url{http://www.et-gw.eu}} (ET), and \textit{LISA}\footnote{\url{https://lisa.nasa.gov}}, will be able to detect IMBH-SBH binaries of different masses (up to $\approx 100-1000\msun$, $\approx 10^3-10^4\msun$ and $\gtrsim 10^4\msun$, respectively). Recently, \citet*{fragk18} investigated the overall IMRI rate across cosmic time from a population of primordial GCs in a Milky Way-like galaxy. They showed that the largest contribution to the rate is due to IMBHs in the more massive clusters. Indeed, when IMBH-SBH binaries merge as a consequence of GW emission, the product is imparted a GW recoil kick which may be up to several thousand km s$^{-1}$ times $\eta^2$, depending on the symmetric mass ratio $\eta=m_1m_2/(m_1+m_2)^2$ (where $m_1$ and $m_2$ are the masses of the IMBH and SBH, respectively) and the relative spin geometry, which removes the IMBH from low mass clusters \citep{lou10,lou11,lou12}.  

If not illuminated by a GW IMRI event or electromagnetic emission due to gas accretion, an IMBH would remain invisible. A couple of bright point-like ultra-luminous X-ray sources ($10^{39}\lesssim L_X/\mathrm{erg\ s}^{-1}\lesssim 10^{41}$) may be explained by an accreting IMBH \citep{kaa17}. Besides GWs, tidal disruption events (TDEs) may also provide a definitive proof of the possible presence of an IMBH in the center of a cluster \citep{ramir09,guill14}. A TDE is the dismantling of a passing star by tidal fields in the vicinity of an IMBH. TDEs have been observed in galactic nuclei, as a consequence of the disruption of a star by an SMBH, but the overall rate is still quite uncertain ($\approx 10^{-5}-10^{-4}$ yr$^{-1}$ per galaxy) \citep{sto16,alex17,lawsmith17,graur18}. TDE involving white dwarfs may take place as well, which might have triggered a thermonuclear explosion and might outshine the disk's Eddington limit emission \citep{rossw08,rossw09,macleod16,lawsm17}. In galactic nuclei, the TDE rate may be enhanced due to the presence of an IMBH via ongoing Kozai-Lidov oscillations \citep{che09,che11,lin15}. Recently, \citet{fraglei18} claimed that all TDEs in galaxies with bulges more massive than $\approx 4.15\times 10^{10}$ M$_{\odot}$ would remain dark, unless a lower-mass secondary SMBH or IMBH is also present. IMBHs in star clusters may also consume stars passing in their vicinity. In a series of papers, \citet{bau04a,bau04b,bau06} used N-body simulations to study the TDE rate by an IMBH in an isolated GC. Recently, \citet{lin18} observed a TDE event consistent with an IMBH in an off-centre star cluster, at a distance of $\sim 12.5$ kpc from the centre of the host galaxy.

In this paper, we address the question of whether the primordial GC population that formed in galaxies of different sizes can retain their IMBHs, and examine the expected rate of TDEs and IMRIs. We model the evolution of GCs in the Galactic field by following the semi-analytical method outlined in \citet{gne14}, and also including the dynamics of the sub-cluster of IMBHs and SBHs which form in the cluster core, as described in \citet{fragk18}. Moreover, we take into account the TDE consumption rate of cluster stars, which helps in consuming the cluster mass across cosmic time.

The paper is organized as follows. In Section \ref{sect:gcev}, we present the semi-analytical method we use to evolve the primordial GC population over a Hubble time. In Section \ref{sect:galmod}, we report the cosmological scaling relations used to construct our host galaxy and GC samples. In Section \ref{sect:evolgcimbh}, we discuss GC and IMBH evolution in the host galaxy. We describe our inferred TDE rates and GW rates in Section \ref{sect:tde} and Section \ref{sect:gw}, respectively. Finally, in Section \ref{sect:conclusions} we draw our conclusions.

\section{Globular Cluster evolution}
\label{sect:gcev}

\begin{table*}
\caption{Galaxy Models and initial mass fraction in GCs: galaxy type, mass in stars ($M_*$), index of the \citet{ser63} profile ($n$), effective radius ($R_e$), mass of the dark matter halo ($M_{DM}$), virial radius ($R_{vir}$), concentration of the dark matter halo ($c_{DM}$), scale radius of the dark matter halo ($r_{DM}$), initial mass fraction in GCs ($f_{GC,i}$), total final mass in GCs ($M_{GC,tot}$), final number of GCs ($N_{GC,f}$).}
\centering
\begin{tabular}{cccccccccccc}
\hline
Type & $M_*$ (M$_\odot$) & $n$ & $R_e$ (kpc) & $M_{DM}$ (M$_\odot$) & $R_{vir}$ (kpc) & $c_{DM}$ & $r_{DM}$ (kpc) & $f_{GC,i}$ & $M_{GC,tot}$ (M$_\odot$) & $N_{GC,f}$ \\
\hline\hline
Early & 1 $\times 10^{10}$ & $3$   & $1.15$ & $4.5\times 10^{11}$ & $77$   & $14.6$ & $5.27$ & $0.02$ & $1.36\times 10^{7}$ & $26$\\
Early & 1 $\times 10^{10}$ & $4$   & $1.15$ & $4.5\times 10^{11}$ & $77$   & $14.6$ & $5.27$ & $0.02$ & $1.53\times 10^{7}$ & $30$\\
Early & 1 $\times 10^{10}$ & $5$   & $1.15$ & $4.5\times 10^{11}$ & $77$   & $14.6$ & $5.27$ & $0.02$ & $1.66\times 10^{7}$ & $35$\\
Early & 1 $\times 10^{10}$ & $6$   & $1.15$ & $4.5\times 10^{11}$ & $77$   & $14.6$ & $5.27$ & $0.02$ & $1.76\times 10^{7}$ & $37$\\
Early & 5 $\times 10^{10}$ & $3$   & $2.82$ & $1.6\times 10^{12}$ & $188$  & $12.4$ & $15.2$ & $0.01$ & $5.41\times 10^{7}$ & $132$\\
Early & 5 $\times 10^{10}$ & $4$   & $2.82$ & $1.6\times 10^{12}$ & $188$  & $12.4$ & $15.2$ & $0.01$ & $5.61\times 10^{7}$ & $141$\\
Early & 5 $\times 10^{10}$ & $5$   & $2.82$ & $1.6\times 10^{12}$ & $188$  & $12.4$ & $15.2$ & $0.01$ & $5.74\times 10^{7}$ & $142$\\
Early & 5 $\times 10^{10}$ & $6$   & $2.82$ & $1.6\times 10^{12}$ & $188$  & $12.4$ & $15.2$ & $0.01$ & $5.82\times 10^{7}$ & $145$\\
Early & 1 $\times 10^{11}$ & $3$   & $4.16$ & $6.7\times 10^{12}$ & $277$  & $10.3$ & $26.9$ & $0.02$ & $2.15\times 10^{8}$ & $558$\\
Early & 1 $\times 10^{11}$ & $4$   & $4.16$ & $6.7\times 10^{12}$ & $277$  & $10.3$ & $26.9$ & $0.02$ & $2.20\times 10^{8}$ & $571$\\
Early & 1 $\times 10^{11}$ & $5$   & $4.16$ & $6.7\times 10^{12}$ & $277$  & $10.3$ & $26.9$ & $0.02$ & $2.23\times 10^{8}$ & $573$\\
Early & 1 $\times 10^{11}$ & $6$   & $4.16$ & $6.7\times 10^{12}$ & $277$  & $10.3$ & $26.9$ & $0.02$ & $2.25\times 10^{8}$ & $583$\\
Late  & 1 $\times 10^{10}$ & $0.5$ & $2.66$ & $4.5\times 10^{11}$ & $177$  & $14.6$ & $12.1$ & $0.018$ & $1.62\times 10^{7}$ & $58$\\
Late  & 1 $\times 10^{10}$ & $1$   & $2.66$ & $4.5\times 10^{11}$ & $177$  & $14.6$ & $12.1$ & $0.018$ & $1.74\times 10^{7}$ & $56$\\
Late  & 1 $\times 10^{10}$ & $2$   & $2.66$ & $4.5\times 10^{11}$ & $177$  & $14.6$ & $12.1$ & $0.018$ & $1.87\times 10^{7}$ & $56$\\
Late  & 5 $\times 10^{10}$ & $0.5$ & $3.86$ & $1.6\times 10^{12}$ & $257$  & $12.4$ & $20.7$ & $0.012$ & $5.65\times 10^{7}$ & $195$\\
Late  & 5 $\times 10^{10}$ & $1$   & $3.86$ & $1.6\times 10^{12}$ & $257$  & $12.4$ & $20.7$ & $0.012$ & $5.73\times 10^{7}$ & $196$\\
Late  & 5 $\times 10^{10}$ & $2$   & $3.86$ & $1.6\times 10^{12}$ & $257$  & $12.4$ & $20.7$ & $0.012$ & $5.75\times 10^{7}$ & $195$\\
Late  & 1 $\times 10^{11}$ & $0.5$ & $4.75$ & $6.7\times 10^{12}$ & $317$  & $10.3$ & $30.8$ & $0.019$ & $2.24\times 10^{8}$ & $594$\\
Late  & 1 $\times 10^{11}$ & $1$   & $4.75$ & $6.7\times 10^{12}$ & $317$  & $10.3$ & $30.8$ & $0.019$ & $2.30\times 10^{8}$ & $607$\\
Late  & 1 $\times 10^{11}$ & $2$   & $4.75$ & $6.7\times 10^{12}$ & $317$  & $10.3$ & $30.8$ & $0.019$ & $2.34\times 10^{8}$ & $622$\\
\hline
\end{tabular}
\label{tab:models}
\end{table*}

In this section, we briefly report the equations used for evolving the GC population \citep*[for details see][and references therein]{gne14}. We assume that the cluster formation rate was a fixed fraction $f_{\mathrm{GC},i}$ of the overall star formation rate \citep*{fag18}
\begin{equation}
\frac{dM_{\mathrm{GC}}}{dt}=f_{\mathrm{GC},i}\frac{dM_{*}}{dt}\ .
\end{equation}
We describe our choice of $f_{\mathrm{GC},i}$ in Sect. \ref{sect:evolgcimbh}. The initial mass of the clusters is sampled from a power-law distribution
\begin{equation}
\frac{dN_{\mathrm{GC}}}{dM_{\mathrm{GC}}}\propto M_{\mathrm{GC}}^{-\beta},\ \ \ \ M_{\min}<M_{GC}<M_{\max}\ ,
\label{eqn:gcmassini}
\end{equation}
where we adopt $\beta=2$, $M_{\min}=10^4\ \mathrm{M}_{\odot}$ and $M_{\max}=10^7\ \mathrm{M}_{\odot}$. We assume that all GCs formed at redshift $z=3$, and calculate their subsequent evolution for $11.5$ Gyr until the present-day \citep{gne14}. 

We describe the GC mass loss via stellar winds, dynamical ejection of stars through two-body relaxation and the stripping of stars by the galactic field. To account for stellar winds, we estimate the main-sequence lifetime (see Eq.~\ref{eqn:mstime}) for a star of given initial mass and convert the relative mass loss in a range of masses to the mass loss in a range of times, from which we calculate as a function of time the fractional mass-loss rate of a given cluster. We take into account mass loss due to two-body relaxation and stripping by the galactic tidal field according to \citep{pri08}
\begin{equation}
\frac{dM_{GC}}{dt}=-\frac{M_{GC}}{\min(t_{\mathrm{tid}},t_{\mathrm{iso}})}\ ,
\end{equation}
where
\begin{equation}
t_{\mathrm{tid}}(r,M_{GC})\approx 10 \left(\frac{M_{GC}}{2\times 10^5\ \mathrm{M}_{\odot}}\right)^{\alpha}P(r)\ \mathrm{Gyr}
\end{equation}
is the typical tidal disruption time \citep{gie08}, and
\begin{equation}
P(r)=41.4\left(\frac{r}{\mathrm{kpc}}\right)\left(\frac{V_{\mathrm{c}}(r)}{\mathrm{km}\ \mathrm{s}^{-1}}\right)^{-1}
\end{equation}
is the (normalized) rotational period of the cluster orbit and $V_{\mathrm{c}}(r)$ is the circular velocity at a distance $r$ from the galactic centre. We set $\alpha=2/3$ \citep{gie08,gne14}. In the limit of a weak tidal field, the evaporation of stars is controlled by internal dynamical evolution. We describe the typical evaporation time in isolation as a multiple of the half-mass relaxation time \citep{gie11,gne14}
\begin{equation}
t_{\mathrm{iso}}(M)\approx 17 \left(\frac{M_{GC}}{2\times 10^5\ \mathrm{M}_{\odot}}\right)\ \mathrm{Gyr}\ .
\end{equation}

When a cluster arrives in the vicinity of the galactic centre, the tidal forces may be strong enough to dissolve the cluster. We assume that a GC is disrupted when the stellar density at the half-mass radius falls below the mean ambient density in the Galactic field \citep{ant13}
\begin{equation}
\rho_{\mathrm{h}}<\rho_*(r)=\frac{V_{\mathrm{c}}^2(r)}{2\pi G r^2}\ ,
\label{eqn:dens}
\end{equation}
where $\rho_*(r)$ is due to the adopted field stellar mass and the growing mass of the nuclear stellar cluster. Following \citet{gne14}, we adopt the average density at the half-mass radius
\begin{equation}
\rho_{\mathrm{h}}=10^3\frac{\mathrm{M}_{\odot}}{\mathrm{pc}^3}\min\left\{10^2,\max\left[1,\left(\frac{M_{GC}}{2\times 10^5\ \mathrm{M}_{\odot}}\right)^2\right]\right\}\ ,
\label{eqn:rhalfm}
\end{equation}
which limits $\rho_{\mathrm{h}}$ to $10^5\ \mathrm{M}_{\odot}$ pc$^{-3}$ in the most massive clusters, which corresponds roughly to the highest observed half-mass density. The lower limit corresponds to the typical observed density of low-mass Milky Way's GCs, while more massive GCs are expected to be in the expansion phase to fill their Roche lobes, during which $\rho_h \propto M^2$ \citep{gielesetal11}. However, we note that a larger $\rho_h$ at a given cluster mass would imply a more compact cluster, that would be tidally dissolved on longer times by the host galaxy. We also note that in our models, we do not take into account tidal shocks, e.g. due a pericenter passage close to the galactic center. Inclusion of tidal shocks would affect mainly the most massive clusters in the innermost regions of the host galaxy, thus reducing the cluster surviving fraction \citep{pri08,ant13,arc14b}.

We initialize the clusters in a spherical distribution, mapping that of the field stars. We consider clusters to be orbiting on a circular trajectory of radius $r$ and take this radius to be the time-averaged radius of the true, likely eccentric, cluster orbit \citep{gne14}. We consider the effects of dynamical friction on cluster orbits by evolving the radius $r$ of the orbit according to \citep{cha43,bin08}
\begin{equation}
\frac{dr}{dt}=-\frac{r^2}{t_{\mathrm{df}}}\ ,
\label{eqn:dynf}
\end{equation}
where
\begin{equation}
t_{\mathrm{df}}(r,M_{GC})\approx 0.45 \left(\frac{M_{GC}}{10^5\ \mathrm{M}_{\odot}}\right)^{-1}\left(\frac{r}{\mathrm{kpc}}\right)^2\left(\frac{V_{\mathrm{c}}(r)}{\mathrm{km}\ \mathrm{s}^{-1}}\right)\ \mathrm{Gyr}\ .
\label{eqn:dynfric}
\end{equation}
Eccentric orbits have shorter dynamical friction timescales \citep{arc14a}, and increase the mass-loss rate and shorten the GC relaxation time \citep{web14}. Thus, some of the clusters may get disrupted earlier than clusters on circular orbits. Since the primordial distribution of GC eccentricities is not well known, and to keep things simple, we include the effect of the deviation of the cluster's orbit from circular by taking into account a correction factor $f_e=0.5$ in Eq.~\ref{eqn:dynfric}, consistent
with the results of simulations by \citep{jia08}.

\subsection{Stellar mass function}

Following the procedure outlined in \citet{leigh13}, for each GC, we adopt an initial stellar mass function (IMF)
\begin{equation}
f_{\rm m}(m) = \frac{dN}{dm} = {\beta}m^{-\alpha},
\end{equation}
where $N$ is the number of stars with a given stellar mass $m$, and $\alpha$ and $\beta$ are constants. We assume a power-law slope of $\alpha = 2.3$ \citep{salpeter55}. The parameter $\beta$ ensures that the correct total stellar mass is preserved when integrating the IMF. It is determined by normalizing the equation above using the total stellar mass
\begin{equation}
M_{GC} = \int_{M_{\rm min}}^{M_{\rm max}} f_{\rm m}(m)mdm,
\end{equation}
where $M_{\rm min}$ and $M_{\rm max}$ are the minimum and maximum stellar masses, respectively.  Integrating this equation and solving for $\beta$ yields
\begin{equation}
\beta = \frac{M_{GC}(2-\alpha)}{m_{\rm max}^{2-\alpha} - m_{\rm min}^{2-\alpha}}
\end{equation}
Plausible values for the upper and lower mass cut-off at low metallicity are $m_{\rm min} = 0.08$ M$_{\odot}$ and $m_{\rm max} = 150$ M$_{\odot}$ \citep[e.g.][]{dabringhausen12}. We emphasize, however, that our results are insensitive to the choice of upper-mass cut-off.

Finally, for the progenitor lifetimes $\tau_{\rm p}$, we assume that the MS lifetime $\tau_{\rm MS}$($m$) provides a good approximation, provided the progenitor mass is $m \le 18$ M$_{\odot}$. This seems justified because the MS lifetimes of low-mass stars greatly exceed that of every other evolutionary phase, typically by several orders of magnitude \citep[e.g.][]{clayton68,iben91,maeder09}. Note that we are ignoring any metallicity dependence in the MS lifetime, since metallicity should only weakly affect it. For the MS lifetime, we assume \citep{hansen94}
\begin{equation}
\tau_{\rm MS}(m) = \tau_{\rm 0}\Big( \frac{m}{\msun} \Big)^{-2.5}
\label{eqn:mstime}
\end{equation}
with $\tau_{\rm 0} = 10^{10}$ yr. For progenitor masses $m > 18$ M$_{\odot}$, we impose a fixed total lifetime of 7 Myr (note that equation 7 yields the same MS lifetime of 7 Myr for $m = 18$ M$_{\odot}$). This is in rough agreement with stellar evolution models, which predict a near-constant lifetime for massive stars at low metallicity \citep[e.g.][]{iben91,hurley00,maeder09}. Thus, our final estimate for the total progenitor lifetime is
\begin{equation}
\tau_{\rm p} = \max(\tau_{\rm MS}(m), 7\,{\rm Myr}).
\end{equation}

We assume that MS stars with progenitor masses $< 8$ M$_{\odot}$ \citep[e.g.][]{maeder09} evolve to become WDs, and more massive progenitors evolve to become either NSs or BHs. With this mass cut-off, we can compute the relative numbers of MS stars and WDs at every time-step in our model. This, in turn, is used to estimate the relative rates of TDEs of MS stars and WDs in Section~\ref{sect:tde}.

\section{Galaxy models}
\label{sect:galmod}

The galactic potential within which our GCs orbit is described by a Sersic profile and a dark matter halo. We also consider the nuclear star cluster as it starts forming. For each galaxy model, we fix the stellar mass content $M_*$ and its Sersic index $n$ \citep{ser63}, and then we compute the other relevant parameters of the stellar mass and dark matter profile from cosmological simulations. To compute mass and other relevant profiles in the Sersic model, we adopt the equations presented in \citet{prug97} and \citet{ter05}. We assume that the dark matter halo is described by a \citet{nfw97} profile. A Sersic profile of total mass $M_*$ is described also by the effective radius $R_e$. We adopt the scaling relation between a galaxy size and its mass from \citet{she03}. They found that such relation can be described by a log-normal distribution with mean
\begin{equation}
\log \left(\frac{R_e}{\mathrm{kpc}}\right)=\log b_1+ a_1\log \left(\frac{M_*}{M_\odot}\right)\ ,
\end{equation}
where $a_1=0.56$ and $b_1=2.88\times 10^{-6}$, for early-type galaxies, and 
\begin{equation}
\log \left(\frac{R_e}{\mathrm{kpc}}\right)=\log c_2+ a_2\log \left(\frac{M_*}{M_\odot}\right)+(b_2-a_2)\log \left(1+\frac{M_*}{M_0}\right)\ ,
\end{equation}
where $a_2=0.14$, $b_2=0.39$, $c_2=0.1$, $M_0=3.98\times 10^{10}$ M$_{\odot}$, for late-type galaxies, and dispersion $\sigma\approx 0.4$.

To link the stellar content of a galaxy to its halo, we assume the following scaling relation between the mass and dark matter content ($M_{DM}$) of a given galaxy \citep{guo10,mos10}
\begin{equation}
M_*=2d_3 M_{DM}\left[\left(\frac{M_{DM}}{M_1}\right)^{-a_3}+\left(\frac{M_{DM}}{M_1}\right)^{b_3}\right]^{-1}
\end{equation}
$a_3=1.068$, $b_3=0.611$, $d_3=0.02817$, $M_1=10^{11.9}$ M$_{\odot}$, and scatter $\sigma_{\log M_*}=0.15$. The virial size $R_{vir}$ of a galaxy scales with $R_e$ as \citep{kra13,som18}
\begin{equation}
R_e=0.015\ R_{vir}\ ,
\end{equation}
with scatter $0.2$ dex. Finally, the concentration of the dark halo, defined as $c_{DM}=R_{vir}/r_{DM}$, where $r_{DM}$ is the scale radius of the halo, scales with the mass of the halo as \citep{bul01}
\begin{equation}
c_{DM}=9\left(\frac{M_{DM}}{1.86 \times 10^{13}\ \mathrm{M}_\odot}\right)^{-0.13}\ ,
\end{equation}
with scatter $\sigma_{\log c_{DM}}=0.14$.

We note that late type galaxies may host disks, which can shock orbiting GCs as they pass through it and enhance the rate of cluster disruption. In this process, stars may gain energy and the cluster binding energy is reduced on average, thus accelerating the escape of stars through evaporation. In general, the inclusion of disk and bulge shocks would accelerate the rate of cluster disruption. In \citet{fragk18}, we found that the typical distance of most surviving clusters is $\sim 2-15$ kpc. In the innermost regions of Milky Way-like galaxies, disc and bulge shocks may disrupt clusters with masses $10^4$-$10^6 \msun$ and half-mass radii $\gtrsim 2-5$ pc \citep{gned97}. For instance, in these galaxies, if shocks were to cause the destruction of all the clusters within $\sim 2$-$5$ kpc, the number of surviving clusters would decrease by a factor of $\sim 15-30$\%.

Table \ref{tab:models} reports all the models used in this work, where host galaxy properties, the initial fraction of galactic mass in GCs, their final total mass and number are listed.

\section{Evolution of the primordial cluster and IMBH populations}
\label{sect:evolgcimbh}

\begin{figure} 
\centering
\includegraphics[scale=0.55]{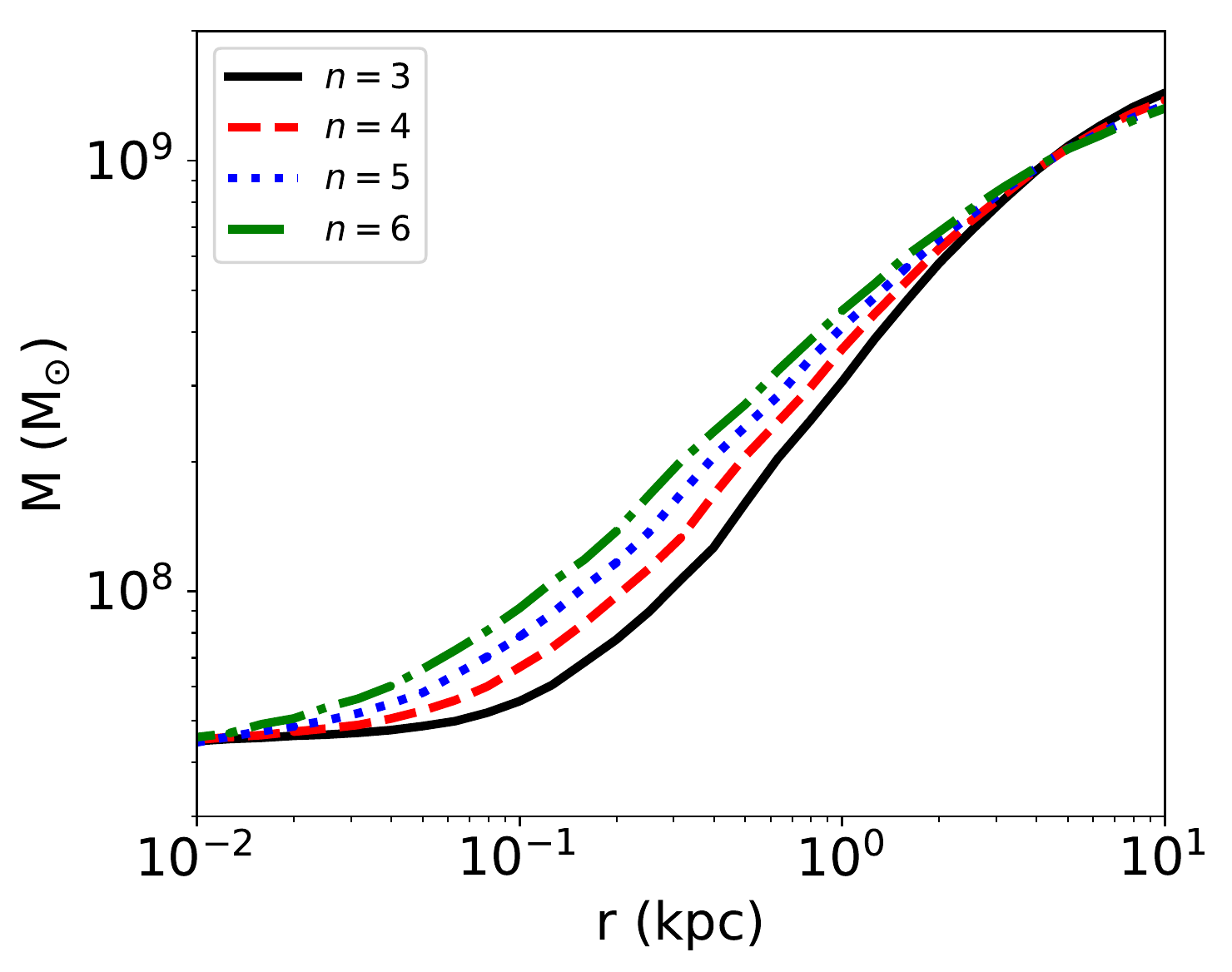}
\includegraphics[scale=0.55]{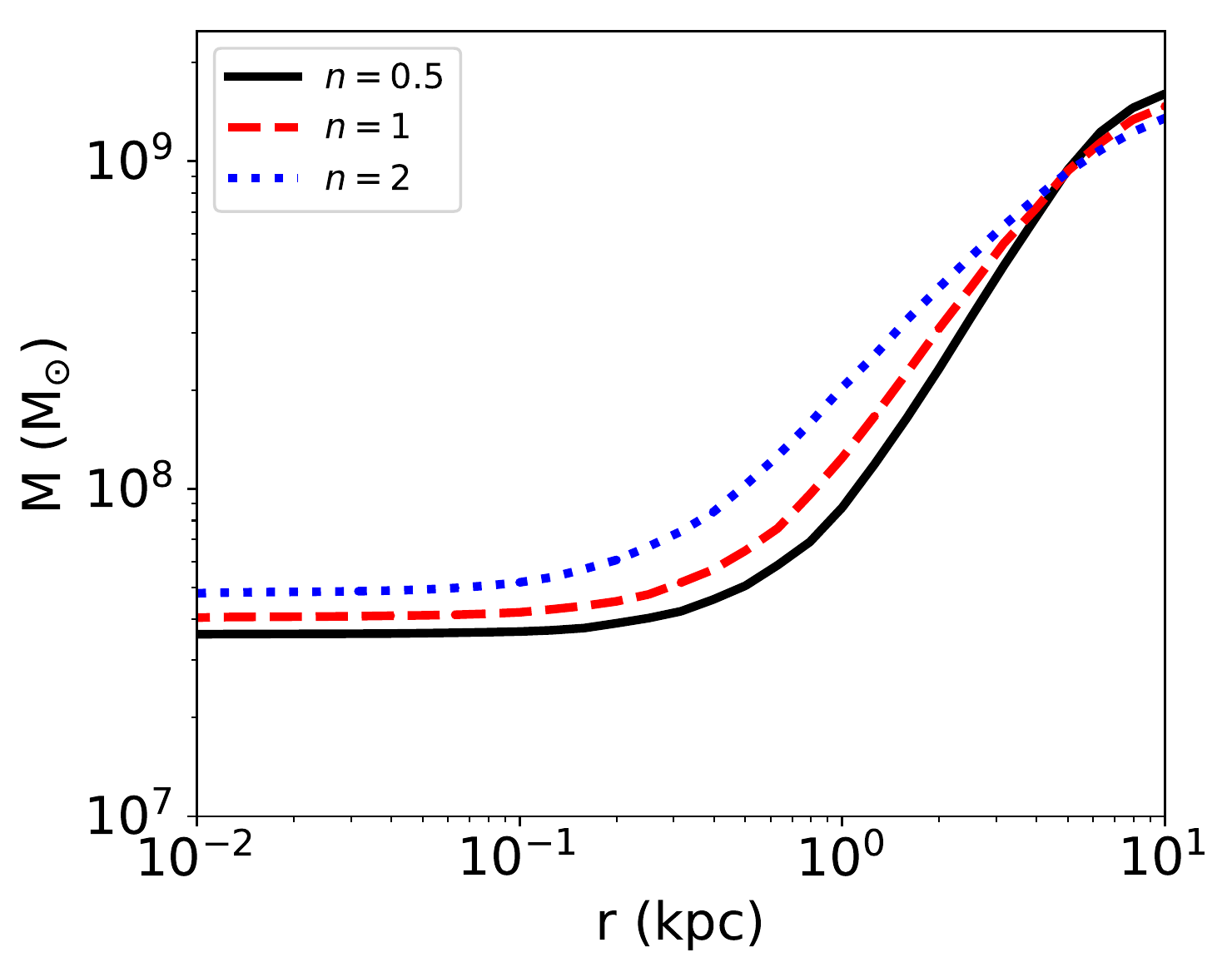}
\caption{Mass accreted onto the galactic nucleus (within $10$ kpc) by disrupted GCs in early type galaxies (top) and late type galaxies (bottom) of stellar mass $M_*=10^{11}\msun$ as a function of the Sersic index. }
\label{fig:nscmass}
\end{figure}

\begin{figure*} 
\centering
\begin{minipage}{20.5cm}
\includegraphics[scale=0.58]{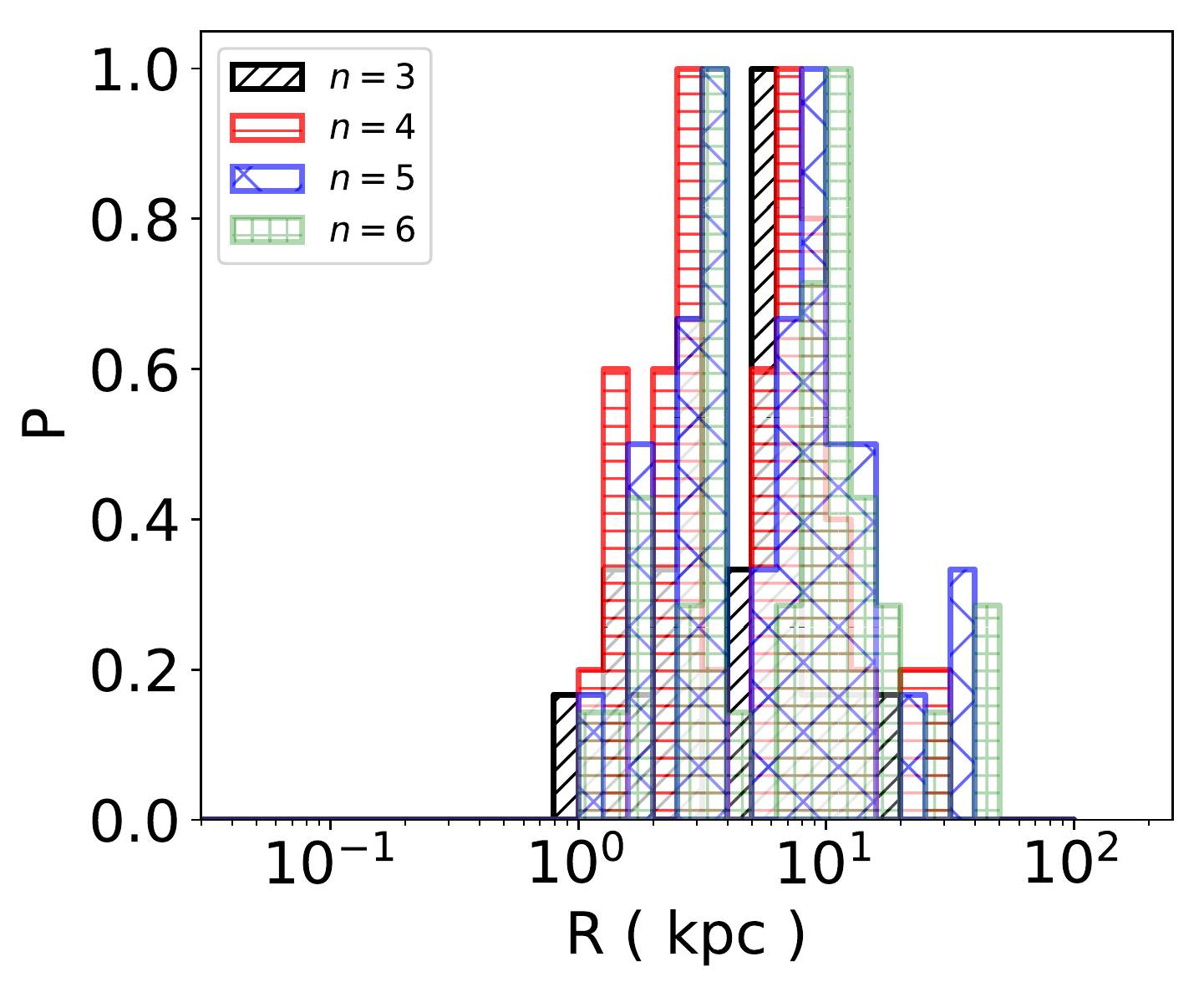}
\includegraphics[scale=0.58]{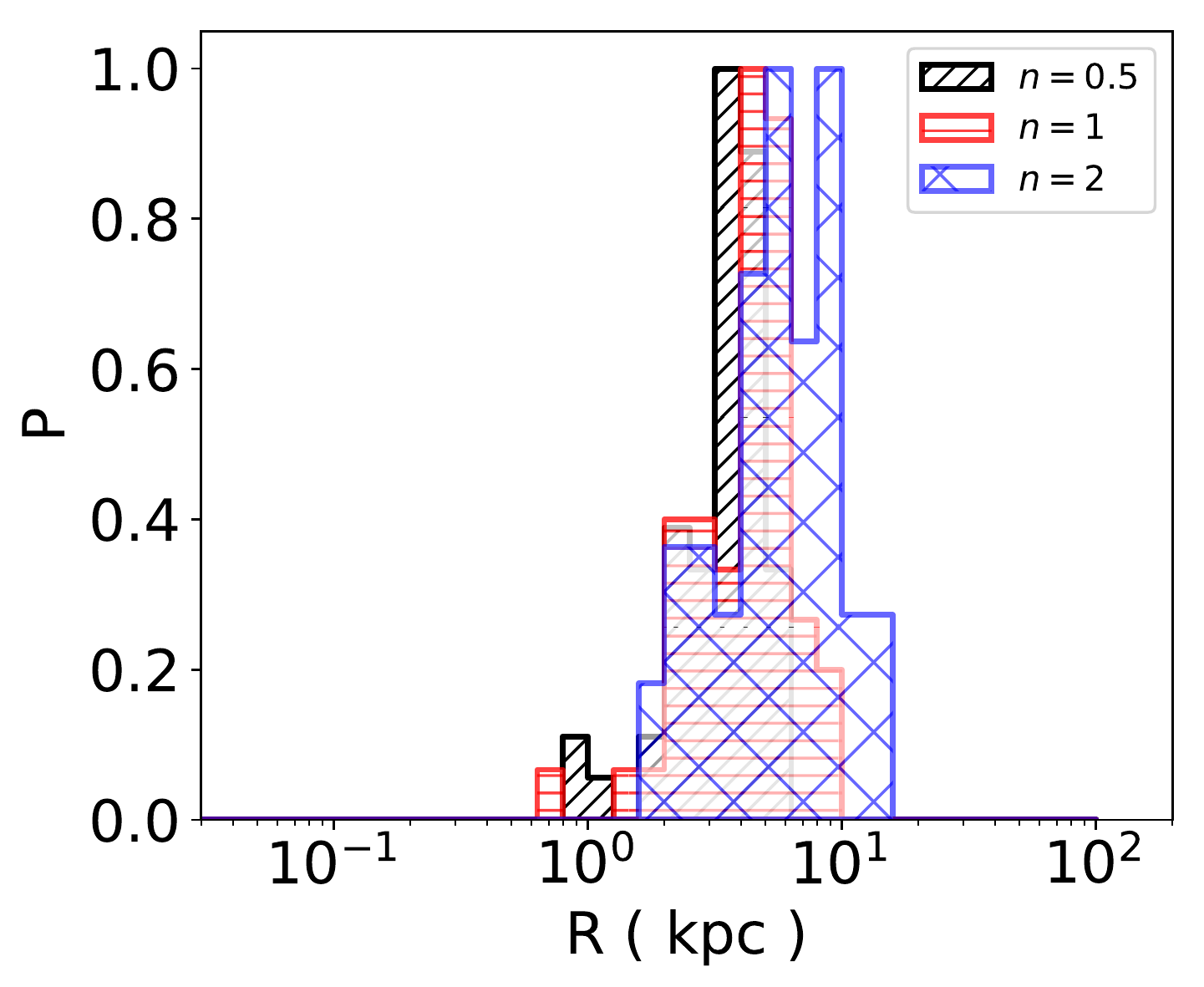}
\end{minipage}
\begin{minipage}{20.5cm}
\includegraphics[scale=0.58]{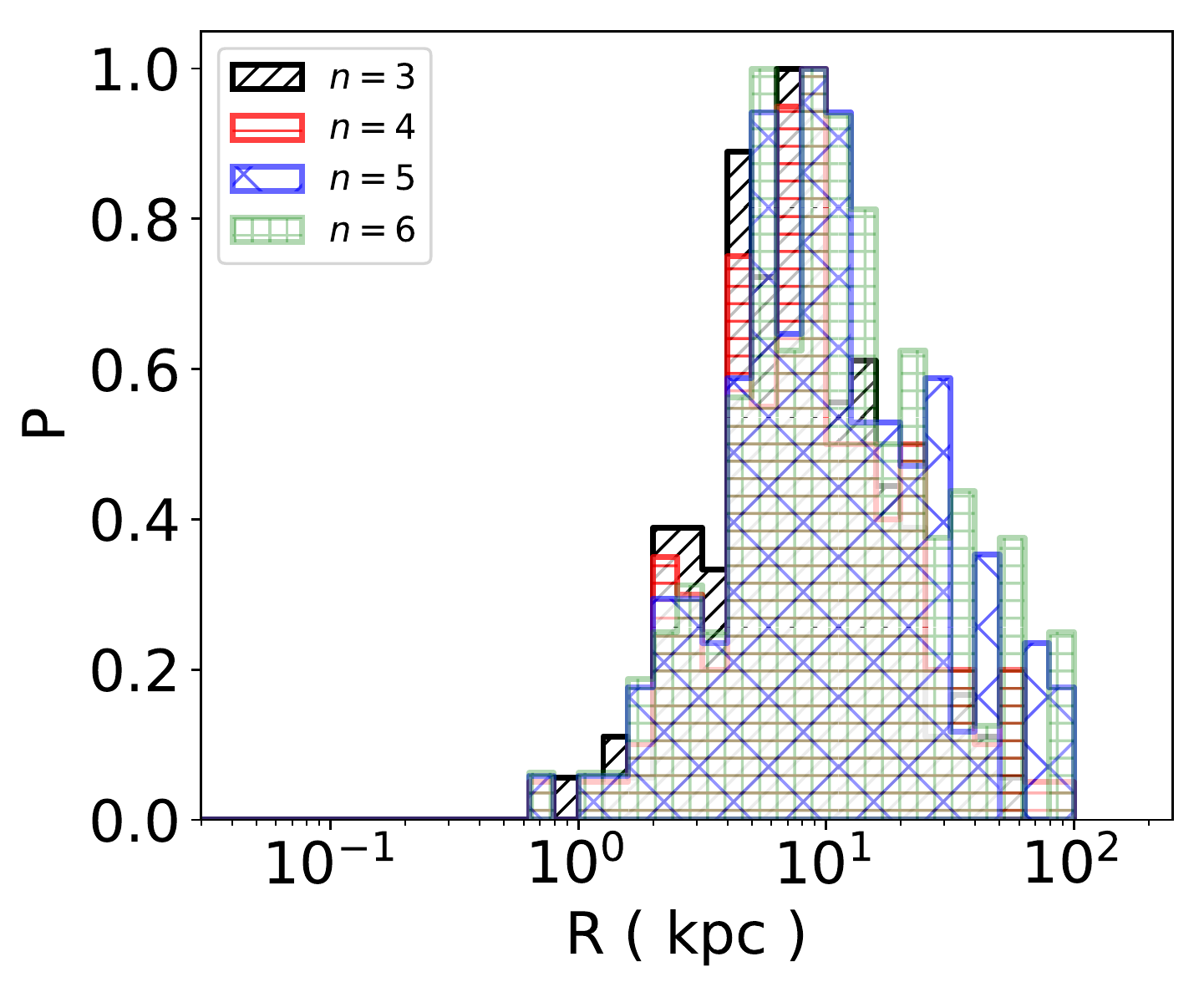}
\includegraphics[scale=0.58]{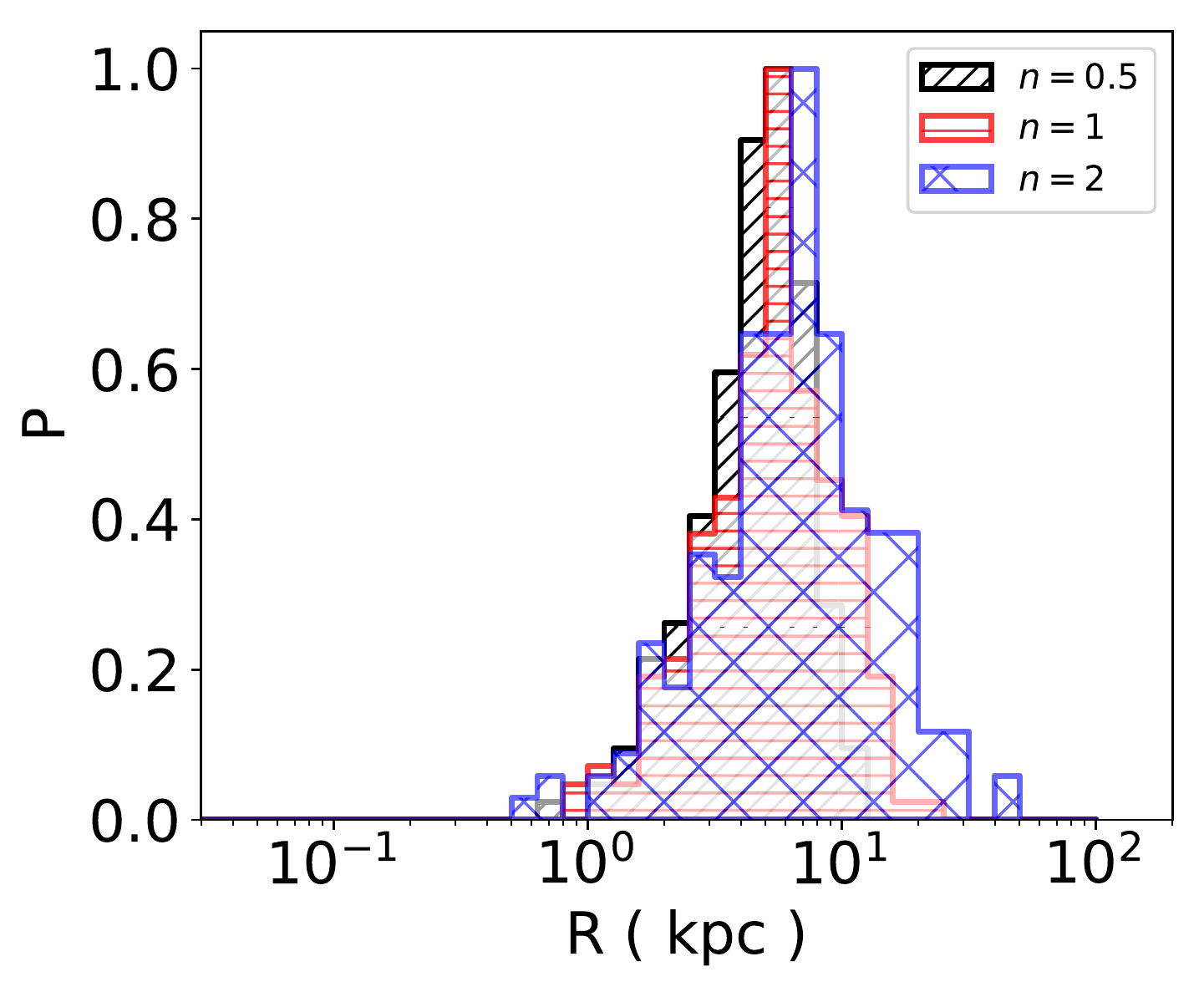}
\end{minipage}
\begin{minipage}{20.5cm}
\includegraphics[scale=0.58]{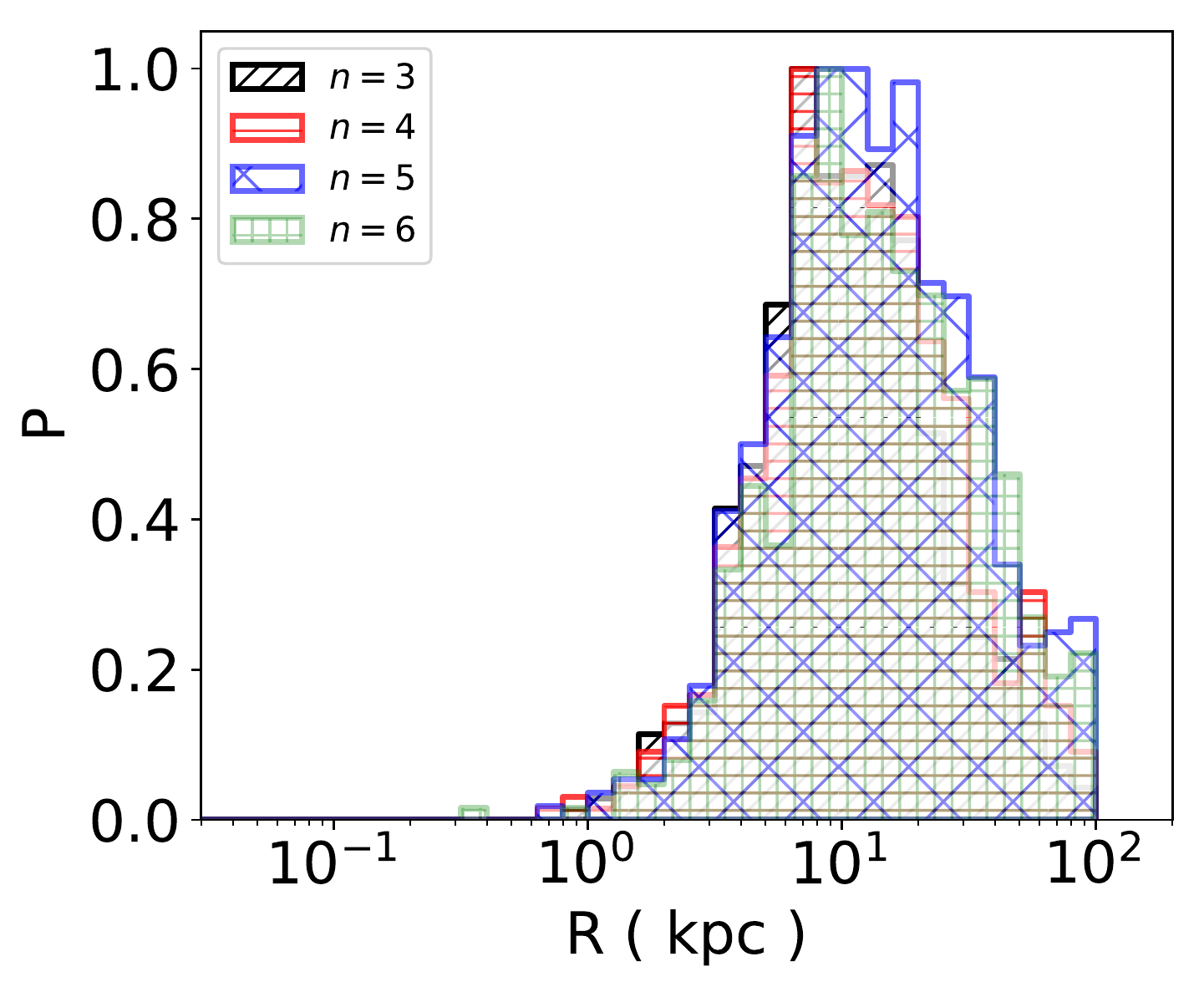}
\includegraphics[scale=0.58]{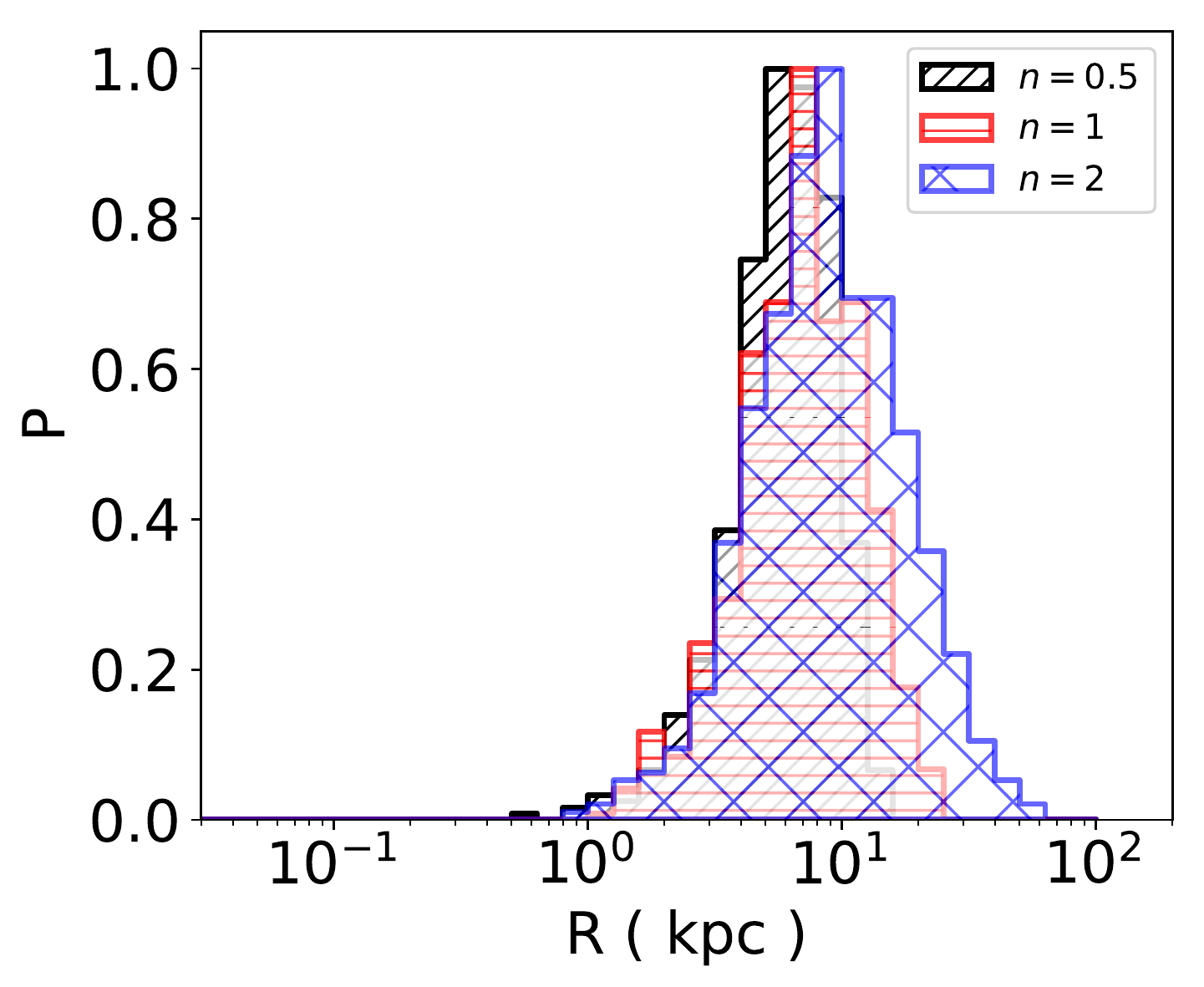}
\end{minipage}
\caption{The radial distribution of surviving GCs (with IMBH in their centres) in early type galaxies (left) and late type galaxies (right) with stellar mass $M_*=1\times 10^{10}\msun$ (top), $M_*=5\times 10^{10}\msun$ (centre) and $M_*=1\times 10^{11}\msun$ (bottom), for different Sersic indexes.}
\label{fig:imbh_final}
\end{figure*}

\begin{figure} 
\centering
\includegraphics[scale=0.58]{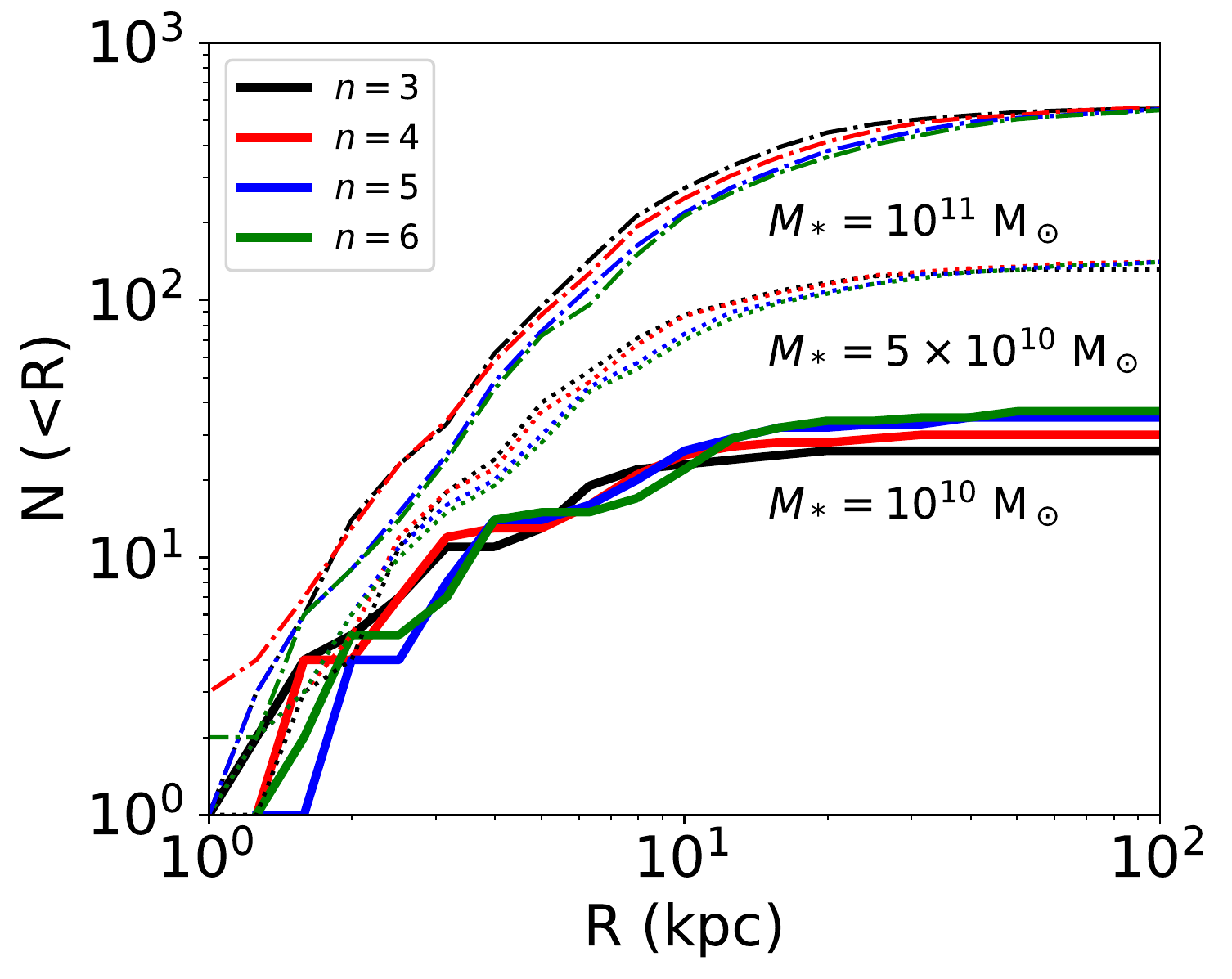}
\includegraphics[scale=0.58]{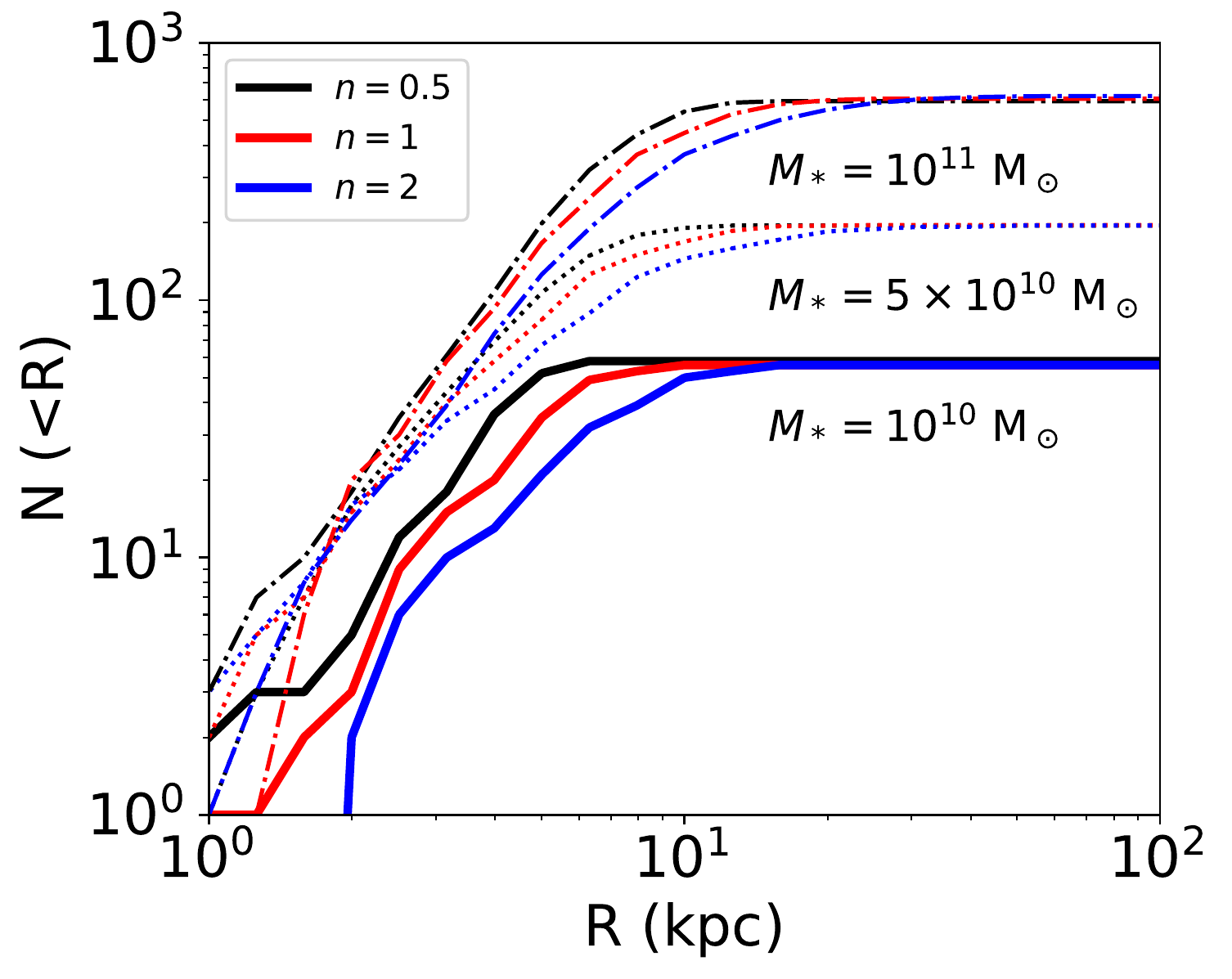}
\caption{The final cumulative radial distribution of IMBHs in star clusters in different early type galaxies (top) and late type galaxies (bottom), and different Sersic indexes.}
\label{fig:imbh_final_rad}
\end{figure}

We evolve the primordial GC populations by means of the equations described in Section \ref{sect:gcev}. The only parameter to be specified in our model is the initial amount of galactic mass in GCs. Unfortunately, the initial cluster mass fraction in GCs $f_{\mathrm{GC},i}$ is poorly understood, but current cluster counts together with radial and mass distributions indicate that it is of the order of a few percent \citep{gne14}. To overcome this problem, we make use of a clear correlation between the present-day mass of the GC population and of the host halo that emerges both from simulations and observations \citep{har13,harris14,harris16,cho18}
\begin{equation}
M_{GC}=3.4\times 10^{-5}\ M_{DM}\ .
\label{eqn:mgcmh}
\end{equation}
with an intrinsic scatter of $0.2$ dex. For each halo mass, we run models with different $f_{GC,i}$ until we satisfy Eq. \eqref{eqn:mgcmh} at present. As shown in Table~\ref{tab:models}, this approach yields GC populations with specific number frequencies (and specific mass frequencies) in agreement with what has been observed for a wide range of galaxy types \citep[e.g.][]{harris99,harris09,harris14,harris16}.

Figure \ref{fig:nscmass} reports the mass accreted onto the galactic nucleus (within $10$ kpc) by disrupted GCs in early type galaxies (top) and late type galaxies (bottom) of stellar mass $M_*=10^{11}\msun$ as a function of the Sersic index. In these galaxies, the typical accreted mass is $\approx 1.5\times 10^9\msun$ and both the galaxy type and Sersic index have only a little impact on the final radial profile of the accreted mass and almost no effect on the total amount. We note that including disk and bulge shocks and cluster eccentric orbits may lead to differences in the final radial profile of the accreted mass \citep{read06,arc14a,petts16}.

While the clusters evolve in the Galactic field, we evolve the IMBH population similarly to \citet[][see also references therein]{fragk18}, whose procedure we briefly summarize here for completeness. In our model, the survival of an IMBH in a GC evolving in a host galaxy is mainly determined by its interactions with the surrounding environment, which is mainly composed of SBHs. However, we note that the physics of IMBH formation and dynamics in GCs is still an active and debated topic, and the detailed composition of its surroundings is unknown. IMBH-SBH interactions commonly happen in the core of the host GC and may kick the IMBH out of the cluster if the GW recoil velocity exceeds the local escape speed. The characteristics of the IMBH-SBH merger events are described by a few parameters
\begin{itemize}
\item the initial fraction of GC mass in IMBHs $f=M_{\rm IMBH}/M_{\rm GC}$
\item the typical timescale $t_{\rm coll}$ between two subsequent IMBH-SBH mergers
\item the slope $\zeta$ of the SBHs mass function
\item the spins $\chi$ of the IMBHs and SBHs
\item the eccentricities $e_{\rm IMBH-SBH}$ of the IMBH-SBH merger event
\end{itemize}
In this paper, we assume for our main model that $f=0.01$, $t_{\rm coll}=50$ Myr \citep{mil02a}, $\zeta=1$, and that both the IMBH and SBHs have zero spin and that $e_{IMBH-SBH}=0$. We discuss how the results depend on different choices for these parameters at the end of the following sections. Apart from IMBH-SBH events, we upgrade the scheme outlined in \citet{fragk18} by self-consistently considering also the effects of ongoing TDEs on the cluster structure. Every time step, the IMBH may grow both because of an IMRI event or because of a TDE event. In a TDE, half of the stellar mass is expected to fall in and be accreted by the IMBH, and half is ejected.  Hence, we increase the IMBH mass by half of the initial stellar mass for each TDE event \citep*{stone13}.

Figures \ref{fig:imbh_final}--\ref{fig:imbh_final_rad} illustrate the radial distribution of IMBHs from the centre of their host galaxy (normalized to the peak value) and the cumulative distribution of GCs surviving until the present time and still hosting an IMBH. As reported in Tab. \ref{tab:models}, the larger the galaxy the larger the number of surviving clusters satisfying Eq. \eqref{eqn:mgcmh}. The main effect of the host galaxy mass is to shift the peak of the distribution to larger distances, from $\approx 5$ kpc in the smallest galaxy considered here to $\approx 12$ kpc in the largest galaxy. We note that GC distributions tend to be broader around the peak of their number distributions in elliptical galaxies relative to spiral galaxies, in which they appear more concentrated around the peak. The Sersic index affects slightly the cluster distribution (the lower $n$ the smaller the peak distance), but it has a negligible effect on the final number of GCs.

\section{Tidal disruption events}
\label{sect:tde}

\begin{figure*} 
\centering
\begin{minipage}{20.5cm}
\includegraphics[scale=0.58]{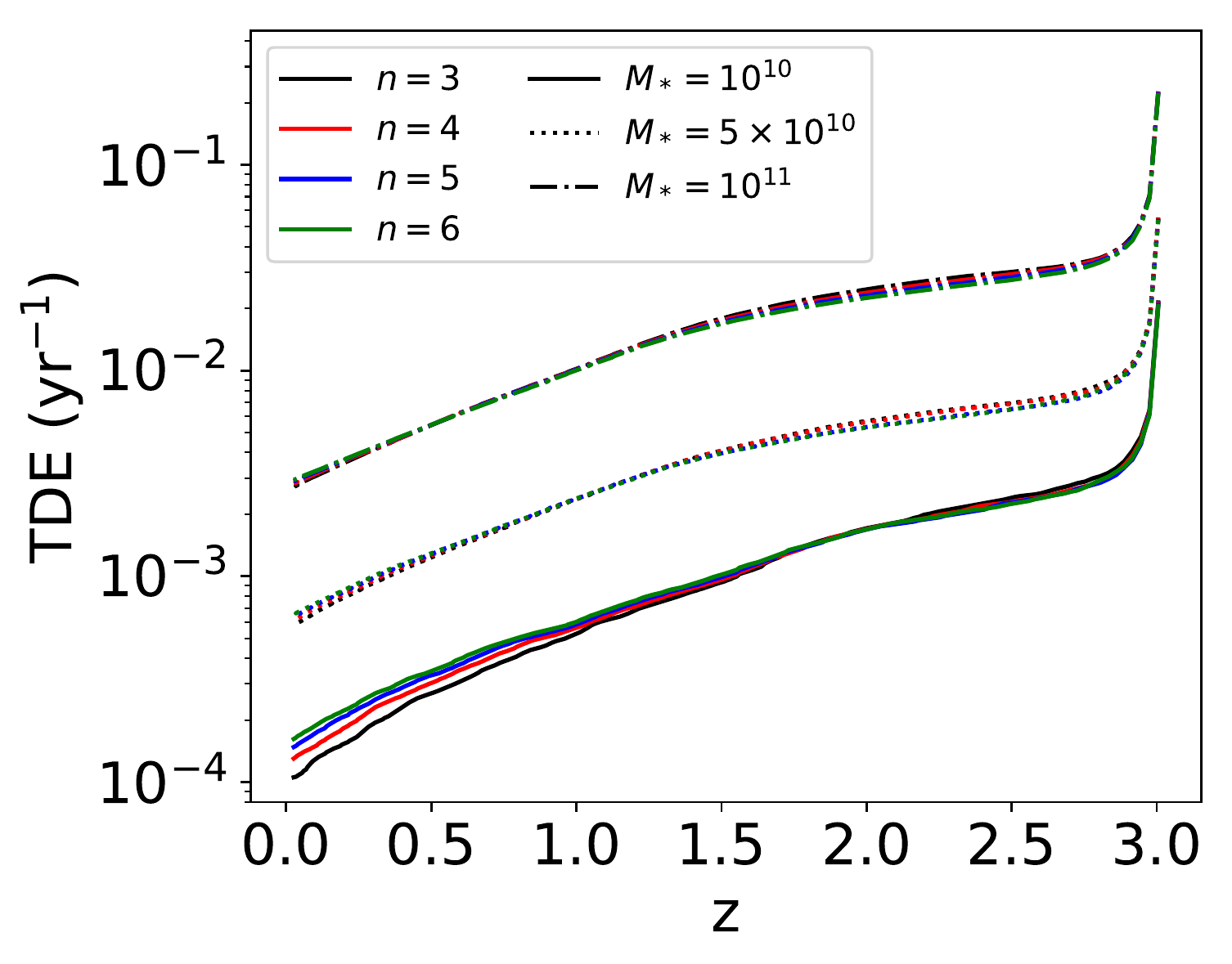}
\includegraphics[scale=0.58]{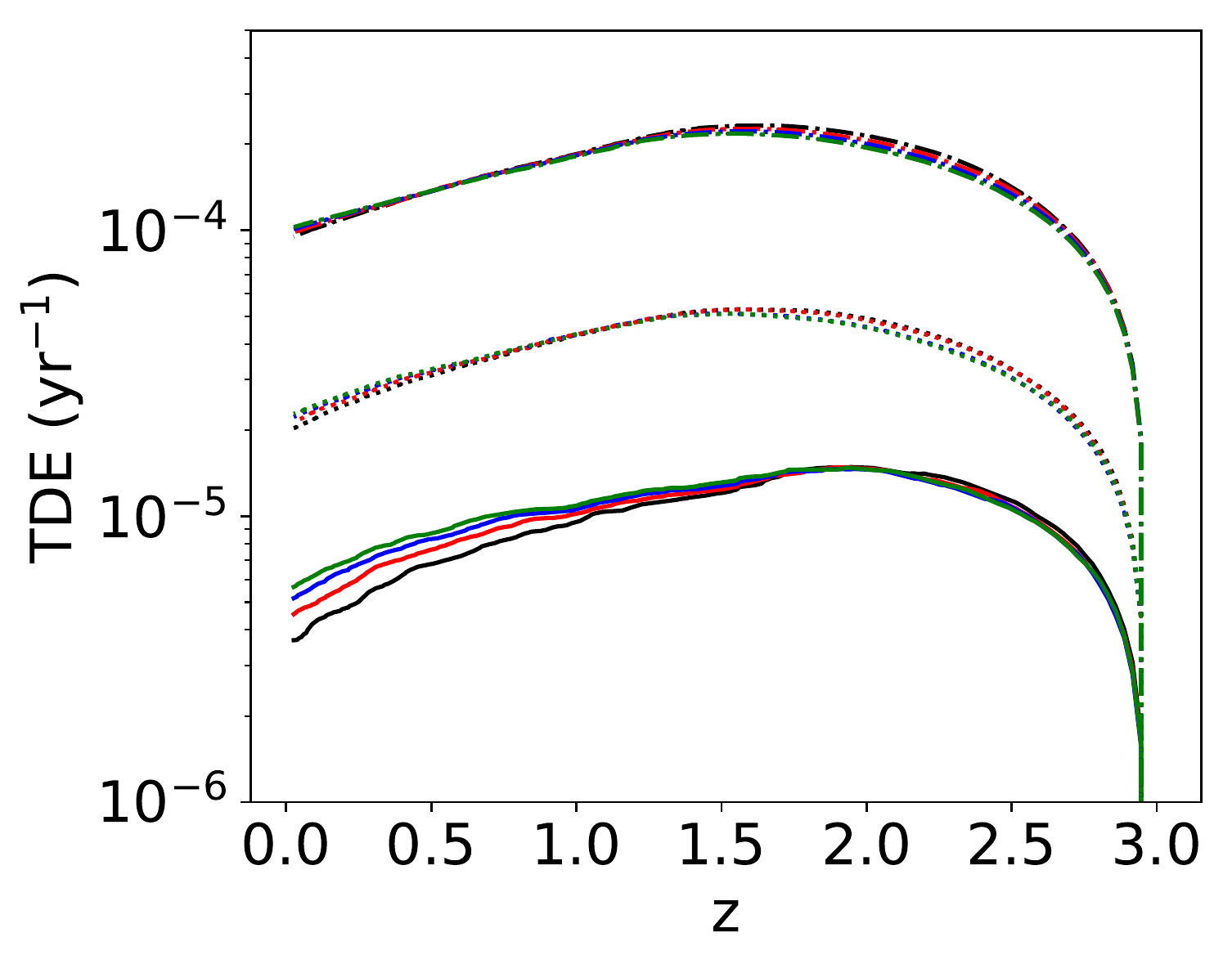}
\end{minipage}
\begin{minipage}{20.5cm}
\includegraphics[scale=0.58]{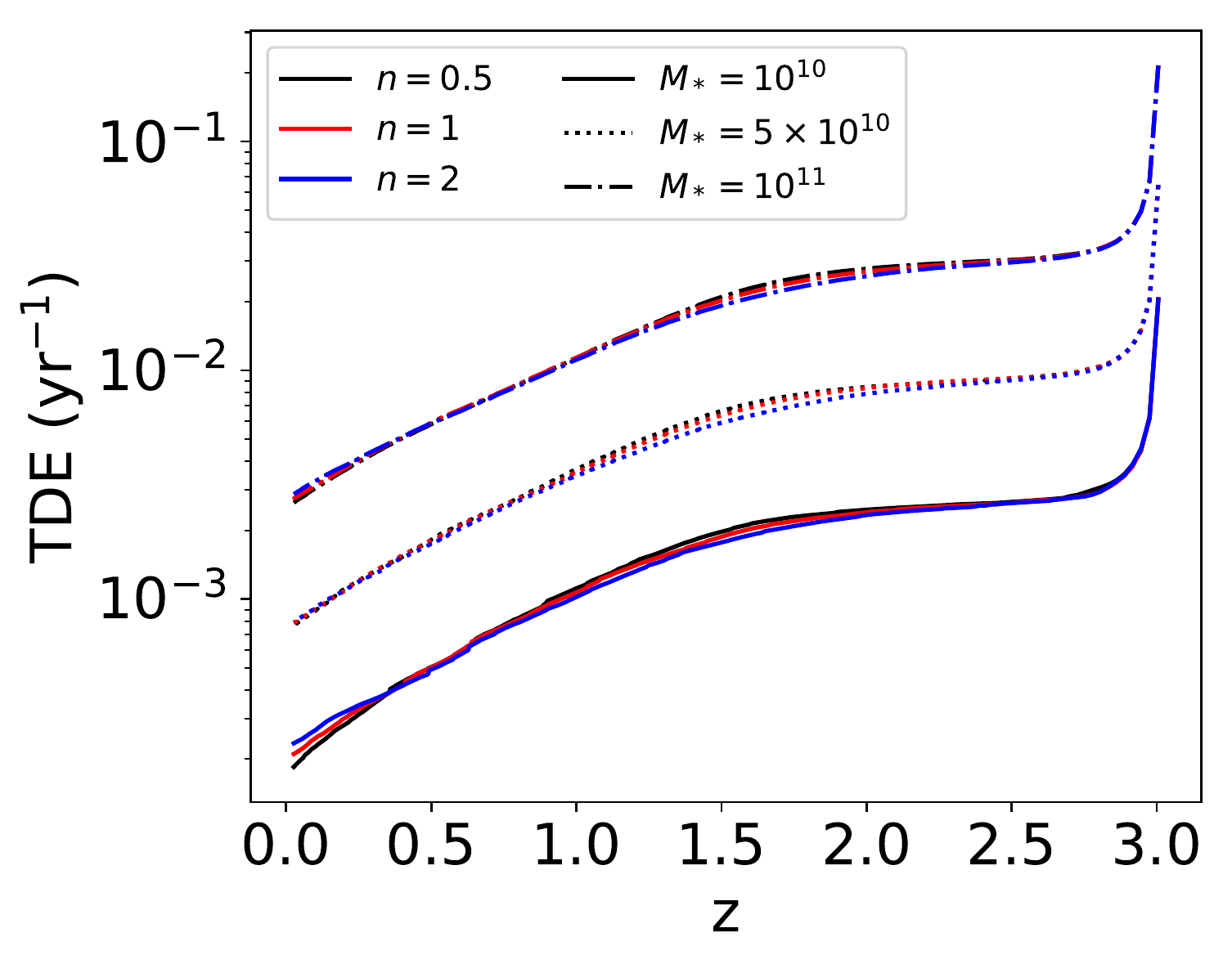}
\includegraphics[scale=0.58]{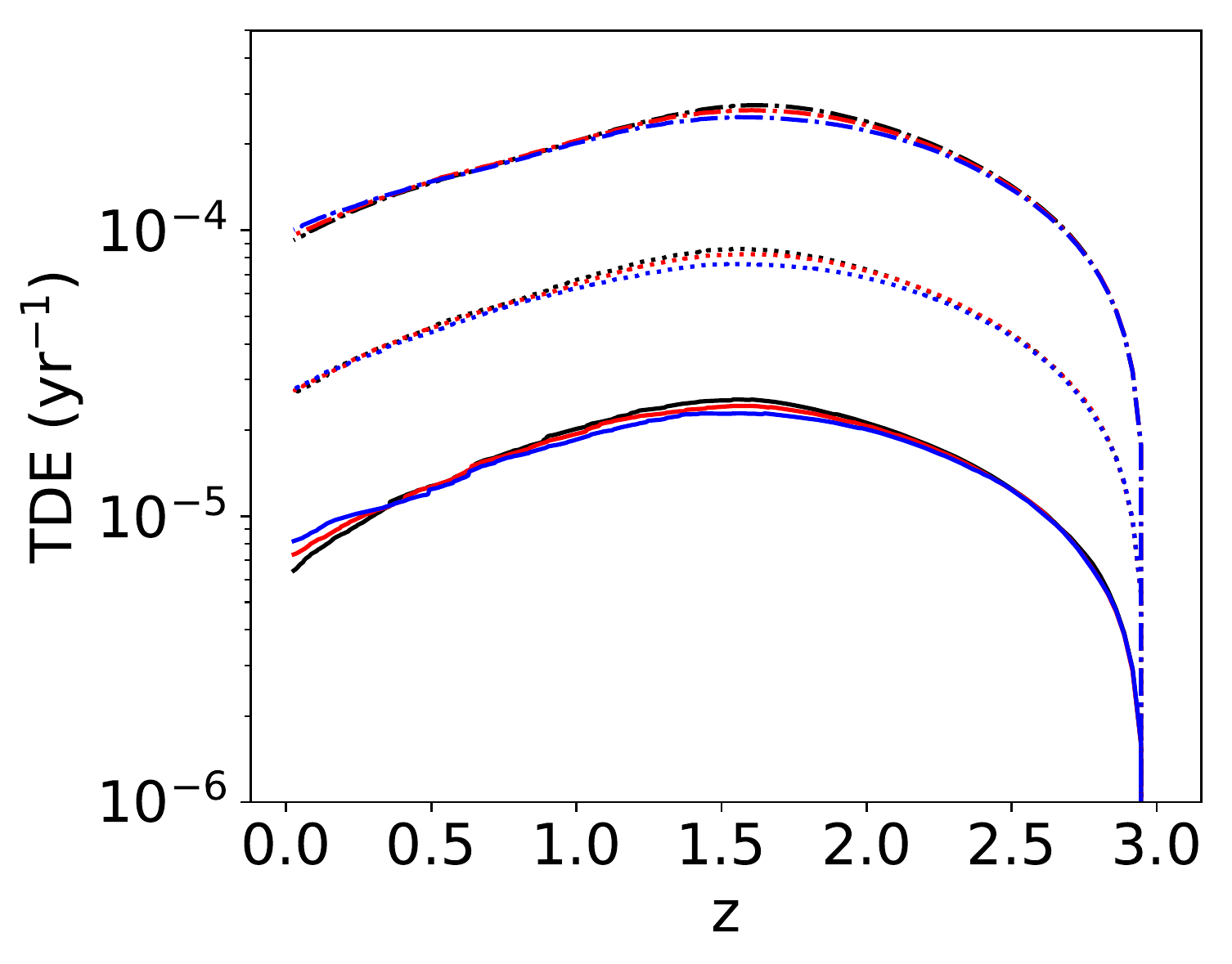}
\end{minipage}
\caption{Comoving TDE rate per galaxy from MS (left) and WD (right) stars for elliptical galaxies (top) and spiral galaxies (bottom) for different Sersic indexes and mass in stars, as a function of redshift $z$. The rates shown are reduced by the initial fraction of GCs hosting IMBHs. TDE rates in elliptical and spiral galaxies are similar to within $10\%$ and MS TDE rates are more common than WD TDE rates by a factor 30 (100) at $z\leq 0.5$ (z= 2).}
\label{fig:tde_singlegal}
\end{figure*}

\begin{figure*} 
\centering
\begin{minipage}{20.5cm}
\includegraphics[scale=0.58]{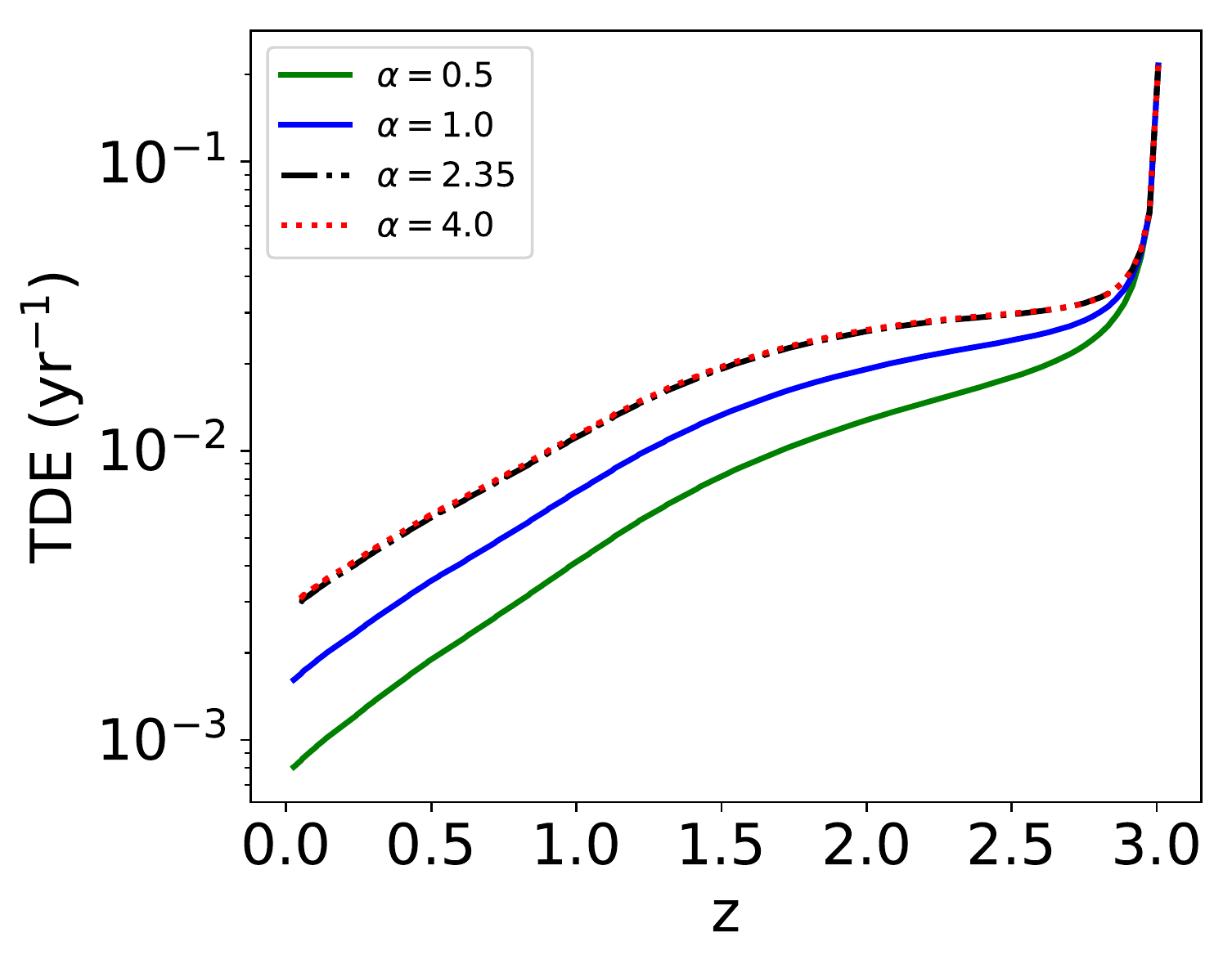}
\includegraphics[scale=0.58]{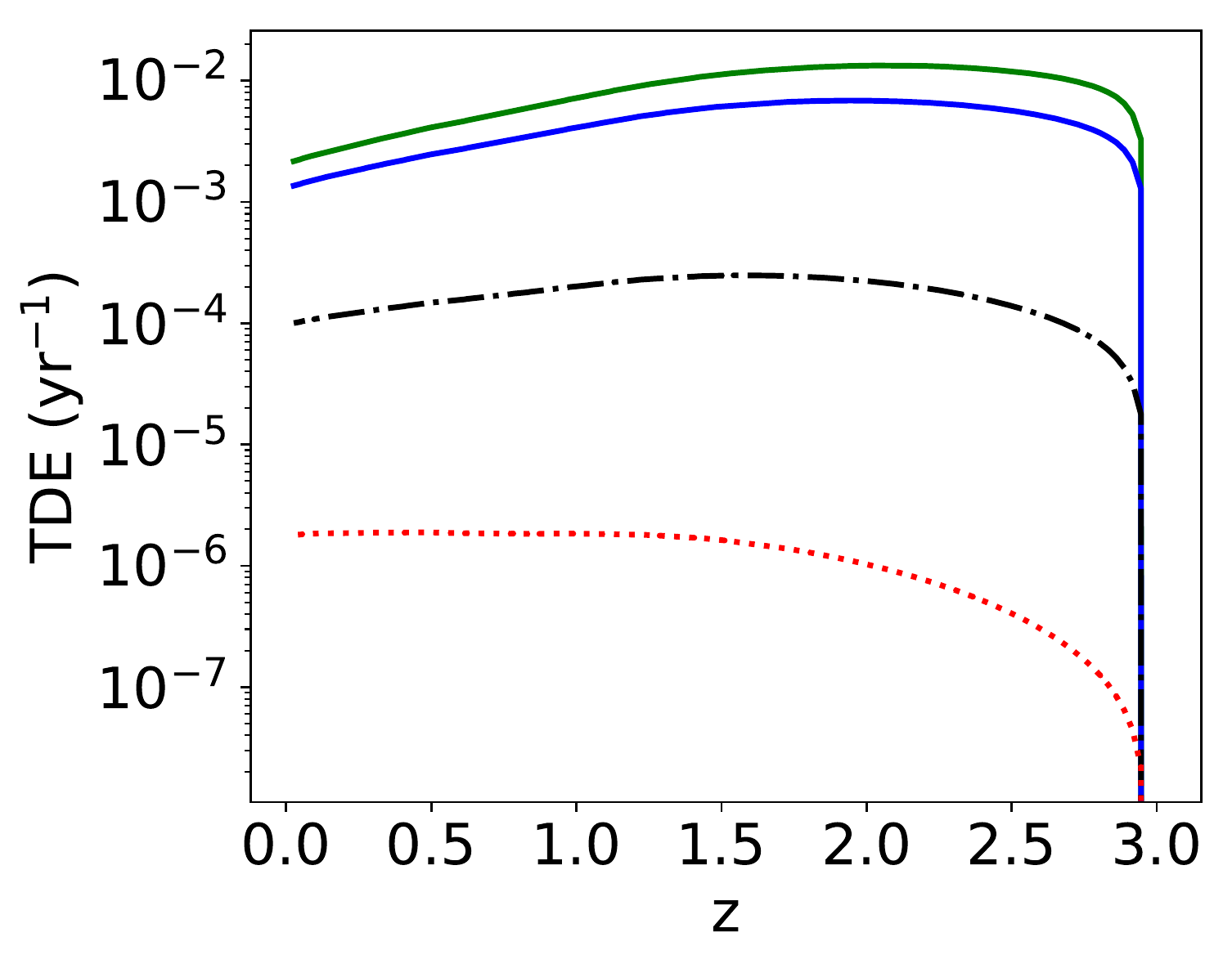}
\end{minipage}
\caption{Comoving TDE rates from MS (left) and WD (right) stars for different mass function exponents $\alpha$ for a spiral galaxy of stellar mass $M_*=10^{11}\msun$ and Sersic index $n=2$.}
\label{fig:tde_alpha}
\end{figure*}

We calculate the TDE rates by IMBHs in GCs in two limiting cases: one in which a density cusp cannot form due to a rapid depletion by TDEs and one in which a density cusp of stars with total mass $M_{IMBH}$ forms around the IMBH. In both cases, we take into account the stars accreated by the IMBH and remove stars from the host GC every time they undergo a TDE.

The tidal disruption of a star by an IMBH occurs when a star on an eccentric orbit has angular momentum smaller than the so-called loss-cone angular momentum \citep{peeb72,bahcall76,ligh77,bau06}
\begin{equation}
J_{LC}\approx\sqrt{2GM_{IMBH}r_t}\ ,
\end{equation}
where
\begin{equation}
r_t=R_*\left(\frac{M_{IMBH}}{m_*}\right)^{1/3}
\end{equation}
is the tidal disruption radius of a star of mass $m_*$ and radius $R_*$. Within the IMBH radius of influence
\begin{equation}
R_{inf}=\frac{GM_{IMBH}}{\sigma_c^2}\ ,
\end{equation}
where $\sigma_c$ is the velocity dispersion in the cluster core, the dynamics is dominated by the IMBH gravitational potential. Nevertheless, stars exchange energy and angular momentum through two-body interactions on the typical relaxation timescale
\begin{equation}\label{eq:T2b}
T_{2b}(a)=\frac{\sigma^3(a)}{\Lambda G^2 \rho(a) m_*}\ ,
\end{equation}
where $\Lambda$ is a dimensionless constant of order of unity (which contains the Coulomb logarithm), $\sigma(a)$ and $\rho(a)$ are the velocity dispersion and density at semi-major axis $a$, respectively. As a consequence of the numerous small-angle two-body deflections, a star with energy $E$ and angular momentum $J$ will diffuse in $E$-$J$ space \citep[for a comprehensive review see][and references therein]{alex17}. Being a random-walk process, the change in angular momentum in a time $t\ll T_{2b}$ is $\delta J\approx (t/T_{2b})^{1/2} J_c$ ($J_c$ is the angular momentum of a circular orbit with the same $a$), which implies a typical timescale
\begin{equation}
T_{2b}(a,e)=\left(\frac{J}{J_c}\right)^2 T_{2b}(a)\ ,
\end{equation}
where $e$ is the eccentricity of the star, to change the angular momentum of order $J$. Thus more eccentric orbits relax more rapidly than circular orbits at the same $a$.

If $\delta J\ll J_{LC}$ (empty loss-cone regime), any star deflected into the loss-cone is disrupted within a dynamical time, while if $\delta J\gg J_{LC}$ (full loss-cone regime) stars may be scattered in and out of the loss-cone on their way from apoapse to periapse \citep{bau04a,bau04b,bau05}. After a rapid initial phase characterized by the wandering of the IMBH through a sea of stars \citep*{chatt02a,chatt02b}, \citet*{stone17} showed that accretion of stars in the full loss-cone regime marks an early stage of black hole accretion, whose rate soon becomes comparable (and then negligible) to the empty loss-cone regime, whicIn ouh dominates at later times. In our calculations we neglect the initial full loss-cone case and assume that the IMBH always accretes at the empty loss-cone rate, thus underestimating the initial TDE rate. Assuming a Plummer model for GCs, we adopt Equation~(18) of \citet{syerulmer99} to compute the empty-loss cone rate as 
\begin{equation}
\Gamma_{\rm TDE} \approx \frac{12}{\pi^2} \frac{G^{1/2} M_{GC}^{1/2}}{R_h^{3/2}} = \frac{24}{3^{1/2}\pi^{3/2}} \sqrt{G\rho_h} \ .
\label{eqn:tdeplu}
\end{equation}
Here $\rho_{h}$ is given as a function of $M_{GC}$ by equation~\eqref{eqn:rhalfm}.

Equation~(\ref{eqn:tdeplu}) shows that the TDE rate $\Gamma$ is independent of the tidal disruption radius $r_t$, and hence independent of the stellar mass or radius. More accurate calculations of the TDE rate show a weak logarithmic dependence on $r_t$ in the empty loss cone regime which modifies the results to within a factor $\sim 2$ \citep[e.g.][]{syerulmer99}. The rate of TDEs of MS stars and WDs can be calculated from equation~(\ref{eqn:tdeplu}) by scaling the result by $N_{MS}/N$ and $N_{WD}/N$, respectively, the number fraction of MSs and WDs with respect to all stars:
\begin{equation}
\Gamma_{MS}=\frac{N_{MS}}{N}\Gamma_{\rm TDE}\quad {\rm and}\quad \Gamma_{WD}=\frac{N_{WD}}{N}\Gamma_{\rm TDE}\,.
\end{equation}
We account for the decrease of the total number of stars due to TDEs. Since we consider self-consistently the evolution of the GC and of the host IMBH, TDEs represent another mass-loss mechanism in addition to those discussed in Sect.~\ref{sect:gcev}. In each timestep, we remove mass (hence stars) from the cluster by the combined effect of stellar winds, internal two-body relaxation, stripping by the host galaxy tidal field and TDEs.

Figure \ref{fig:tde_singlegal} illustrates the TDE rates from MS (left) and WD (right) stars for elliptical galaxies (top) and spiral galaxies (bottom) for different Sersic indexes and galactic masses in stars as a function of redshift $z$. The largest galaxies are expected to have larger TDE rates as a consequence of having the largest numbers of initial and surviving clusters. The rapid increase of the rates at large redshifts is due to the large number of clusters at the beginning of our simulations, some of which (the less massive) rapidly evaporate or get disrupted. Moreover, independently of the galaxy mass, we note a similar decrease in time of the TDE rate. We note that the WD rate increases at high redshift when the WD creation rate dominates, while decreases at later times when the GC disruption rate dominates. The MS TDE rate is $\approx 10^{-4}-10^{-3}$ yr$^{-1}$, while the WD rate is $\approx 30$ times smaller in every galaxy. Both galaxy type and Sersic index do not have a significant effect on the rate. 

Figure~\ref{fig:tde_alpha} shows a comparison of the TDE rate as a function of redshift for different values of the stellar mass function slope $\alpha$.  These mass function slopes are meant to be representative of different time-averaged values, accounting very approximately for the depletion of preferentially low-mass stars due to internal two-body relaxation \citep{leigh12}.  The green and red curves showing $\alpha=0.5$ and $4$ can be interpreted as a rough estimate for the maximum and minimum values for the evolution of the stellar mass function due to two-body relaxation. It is clear from the comparison of these curves that top heavy mass functions imply significantly increased relative rates of TDEs of WDs versus MS stars by up to a factor of 100. In GCs most affected by two-body relaxation, we may even expect the WD TDE rate to surpass the MS TDE rate at redshift $z = 0$, if $dN/dm\propto m^{-0.5}$.

\begin{figure} 
\centering
\includegraphics[scale=0.58]{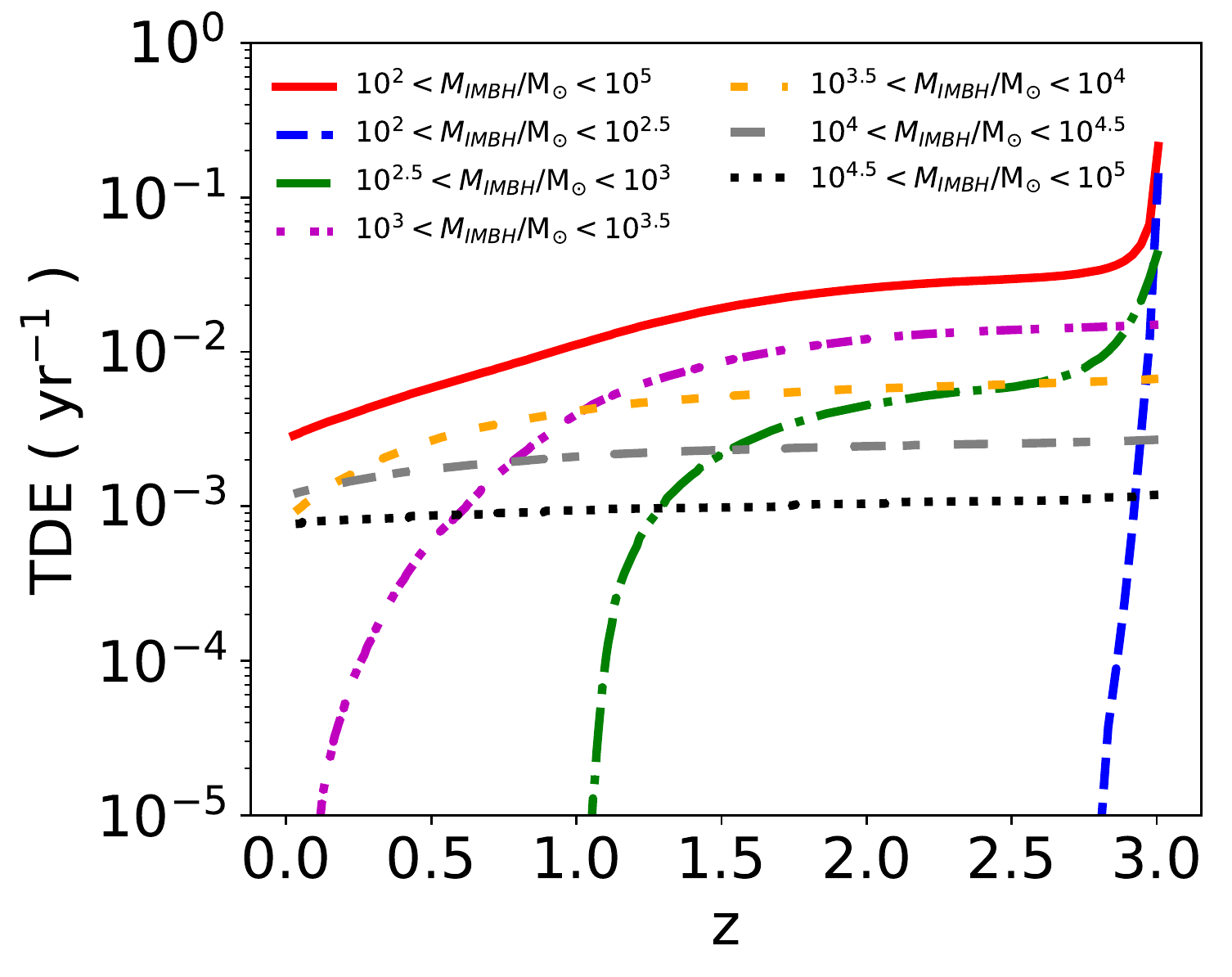}
\caption{Relative contribution of IMBHs of different masses to the comoving TDE rate from MS stars for a spiral galaxy of stellar mass $M_*=10^{11}\msun$ and Sersic index $n=2$.}
\label{fig:tdeimbh}
\end{figure}

\begin{figure} 
\centering
\includegraphics[scale=0.58]{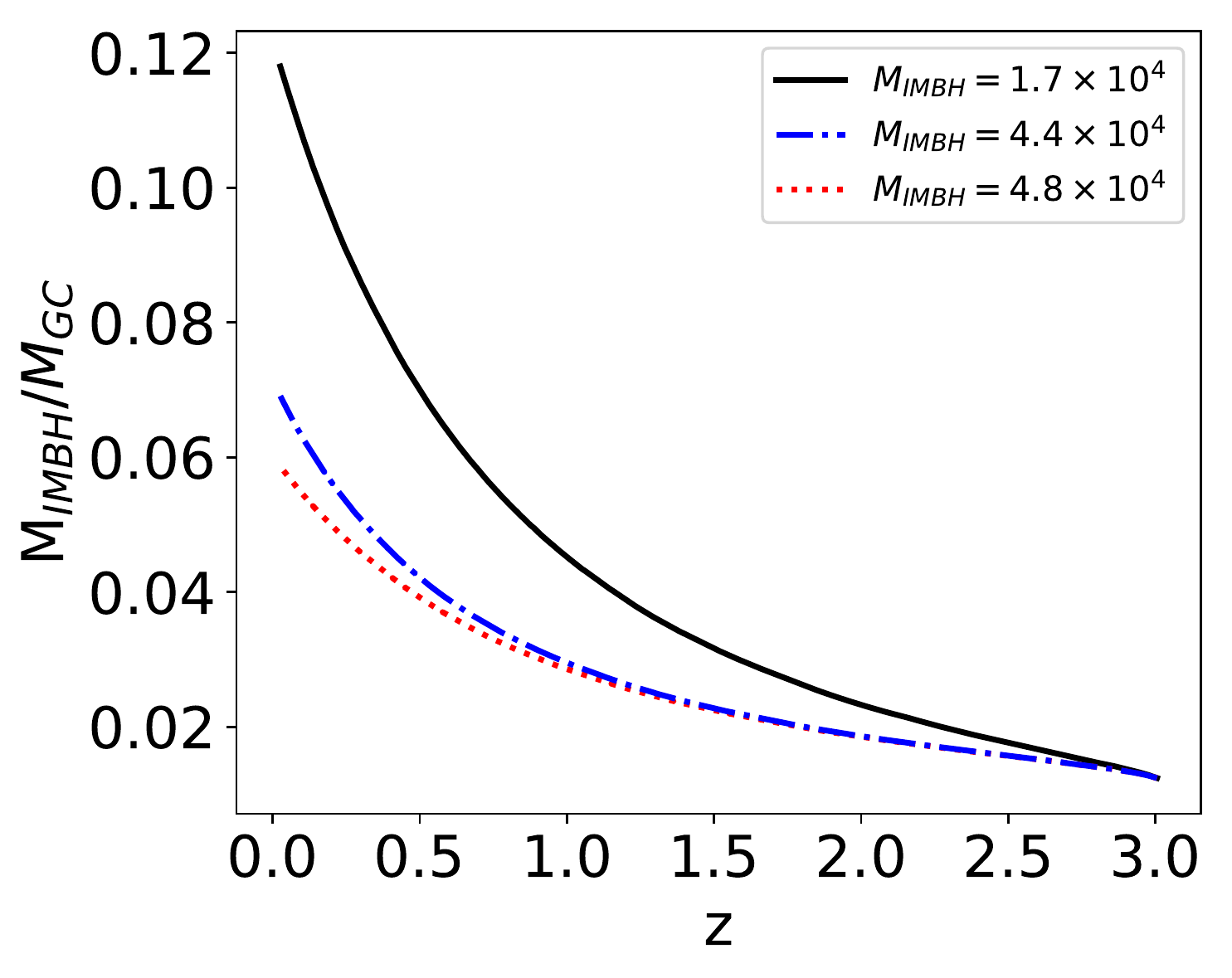}
\caption{Evolution in redshift of the ratio $M_{IMBH}/M_{GC}$, for different IMBH initial masses in GCs of initial mass $\sim 10^6-5\times 10 ^6\msun$. The IMBH mass grows from $1\%$ of the GC mass to $\sim 6-12\%$.}
\label{fig:gcimbh}
\end{figure}

In Fig. \ref{fig:tdeimbh}, we report the relative contributions of IMBHs of different masses to the MS TDE rate for a spiral galaxy of stellar mass $M_*=10^{11}\msun$ and Sersic index $n=2$. IMBHs less massive than $\sim 1000\msun$ produce a significant TDE rate up to $z\gtrsim 1$. Massive IMBHs ($\gtrsim 1000\msun$) have a rate $\sim 10^{-4}-10^{-3}$ yr$^{-1}$, contributing to most of the integrated galactic TDE rate. In Fig. \ref{fig:gcimbh}, we show the evolution in redshift of the ratio $M_{IMBH}/M_{GC}$ for three illustrative cases for the GC initial masses $\sim 10^6-5\times 10 ^6\msun$, and IMBH initial masses $\sim 10^4-5\times 10 ^4\msun$. The IMBH mass grows from $1\%$ of the GC mass to $\sim 6-12\%$, both because the cluster loses mass due to stellar winds and tidal stripping by the galaxy and because the IMBH accretes mass from TDEs. In smaller clusters, the ratio can even grow to larger values.

\begin{figure} 
\centering
\includegraphics[scale=0.58]{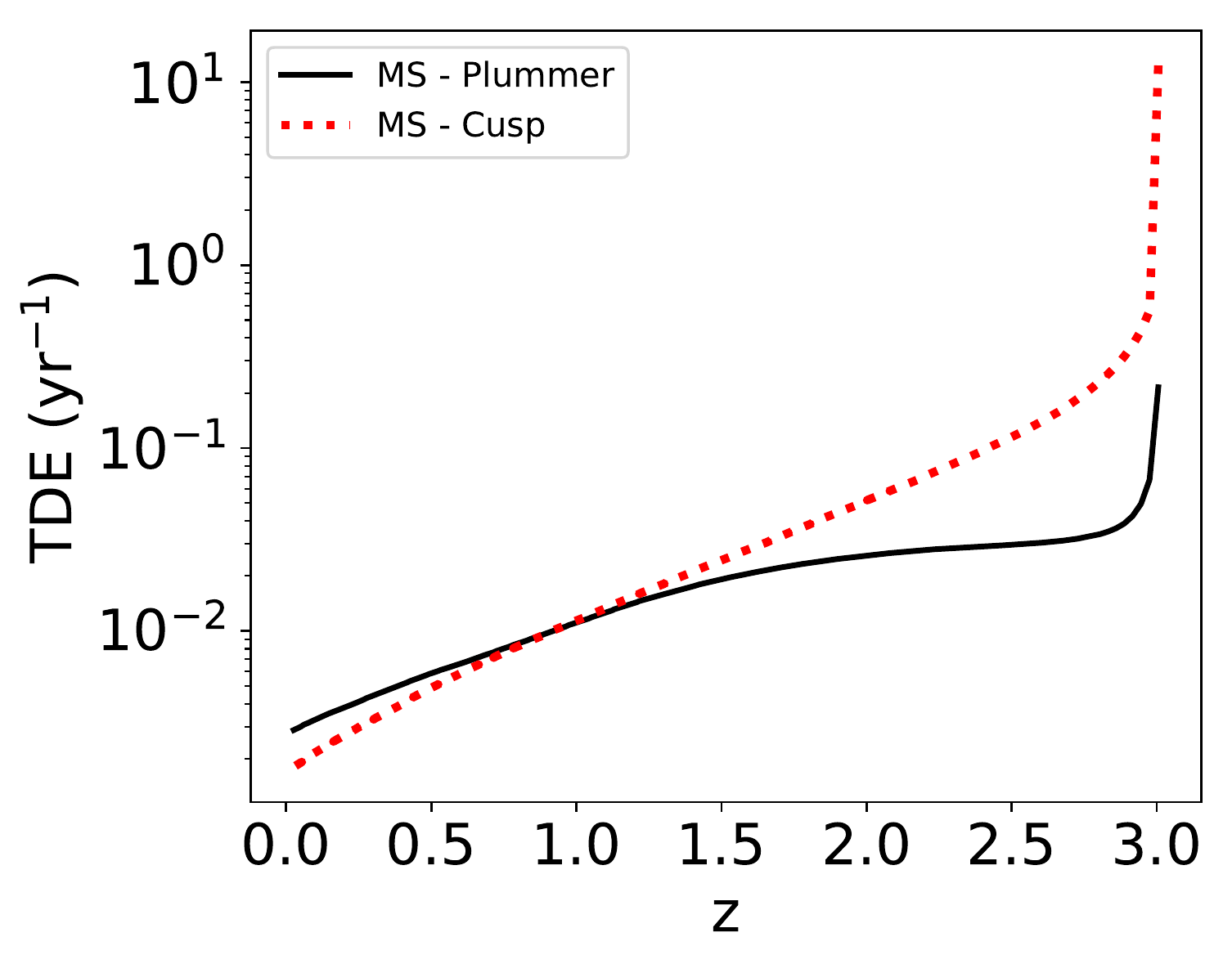}
\caption{Comoving TDE rates from MS stars assuming a cusp profile (Eq. \ref{eqn:tde}) by neglecting TDEs from outside the cusp (dotted line) and for a Plummer profile without a cusp (solid line, Eq. \ref{eqn:tdeplu}) for a spiral galaxy of stellar mass $M_*=10^{11}\msun$ and Sersic index $n=2$.}
\label{fig:plum}
\end{figure}

\begin{figure*} 
\centering
\begin{minipage}{20.5cm}
\includegraphics[scale=0.58]{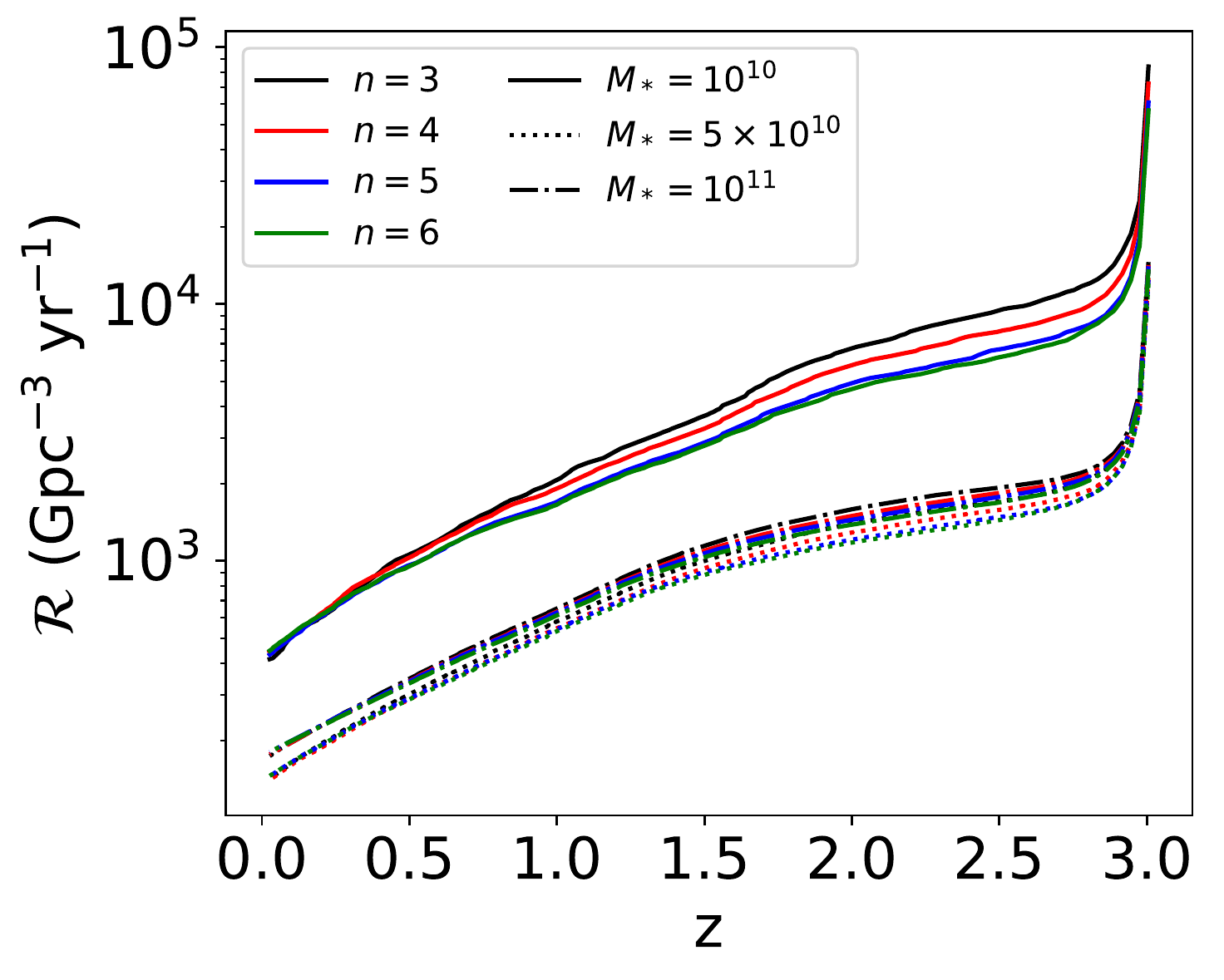}
\includegraphics[scale=0.58]{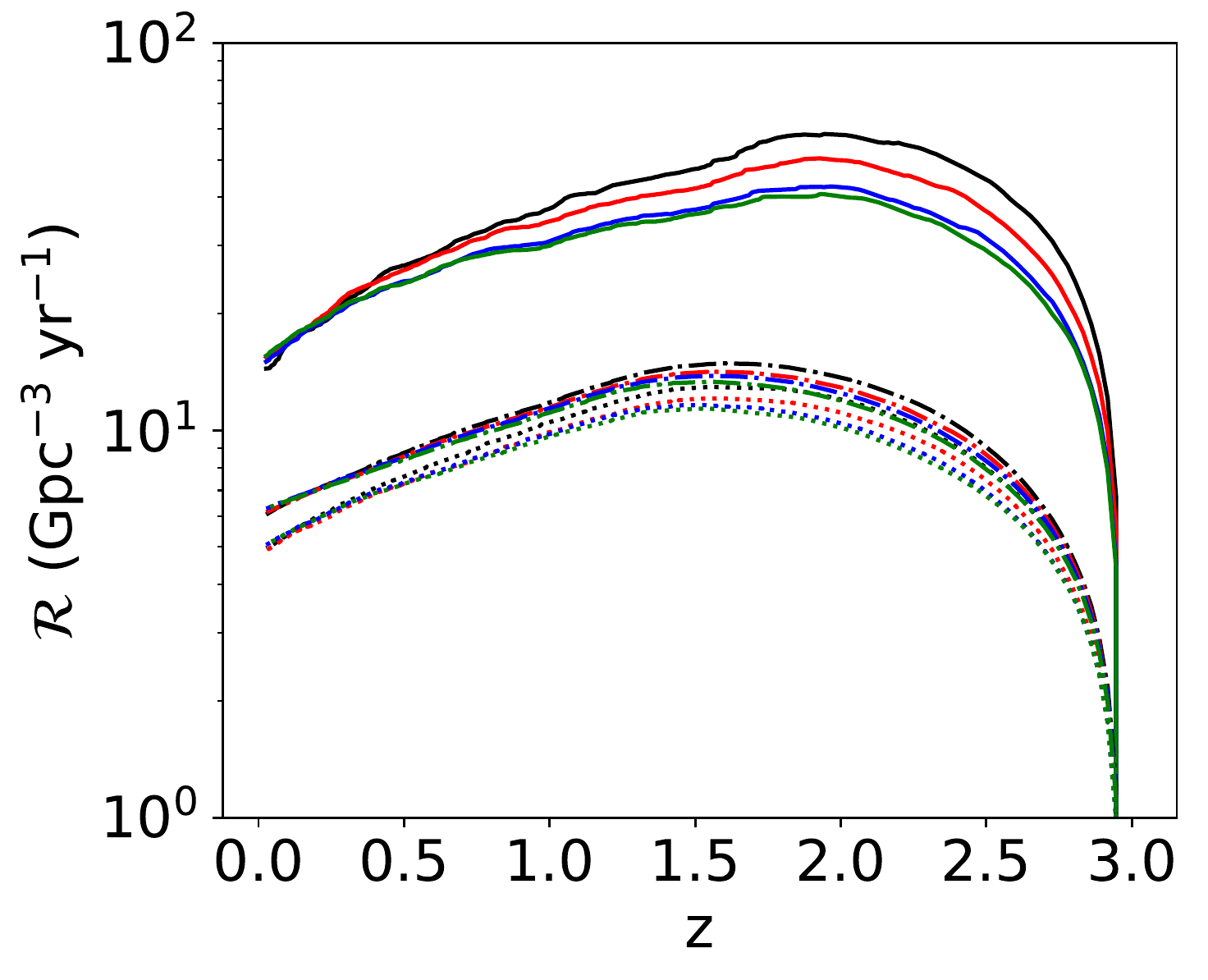}
\end{minipage}
\begin{minipage}{20.5cm}
\includegraphics[scale=0.58]{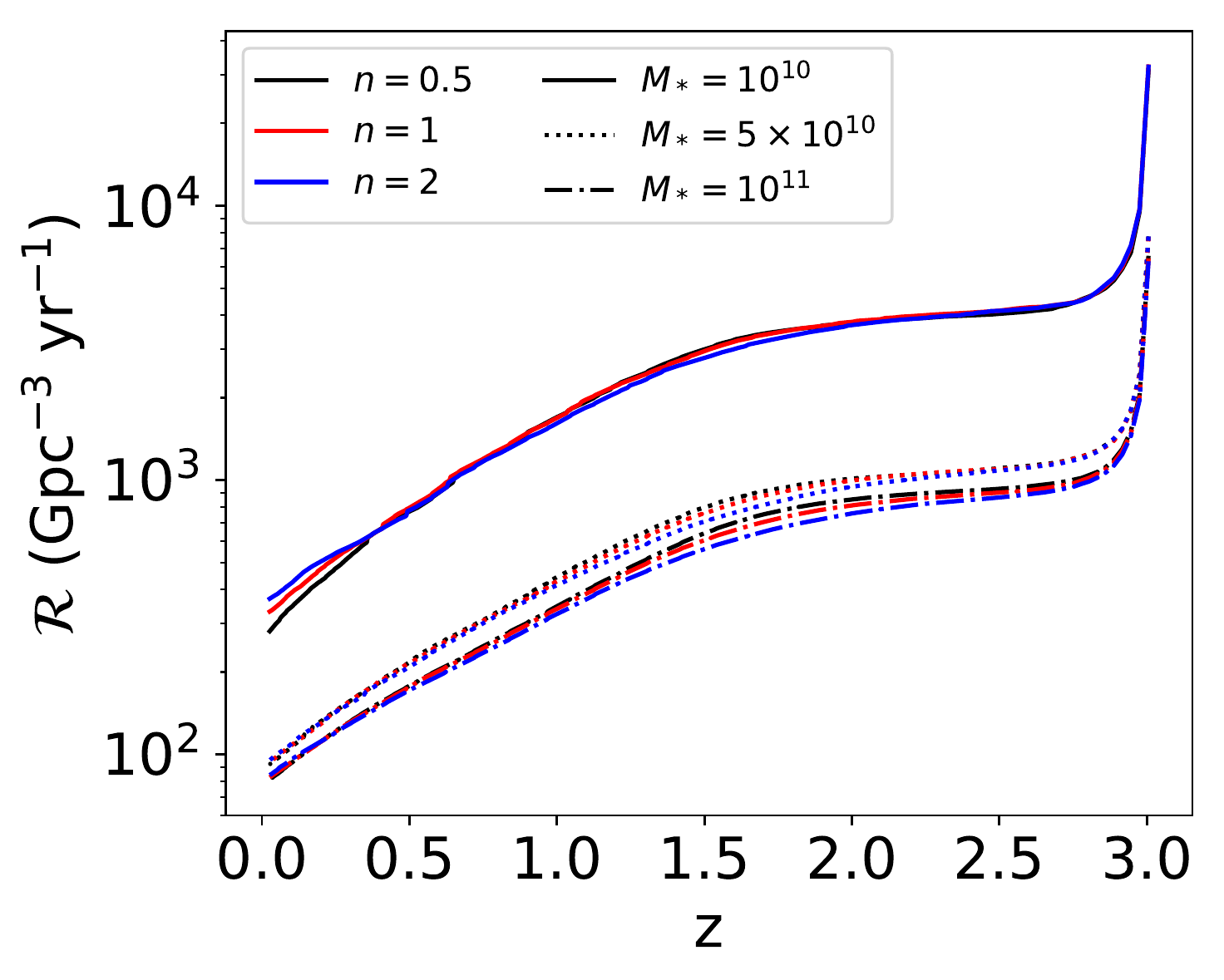}
\includegraphics[scale=0.58]{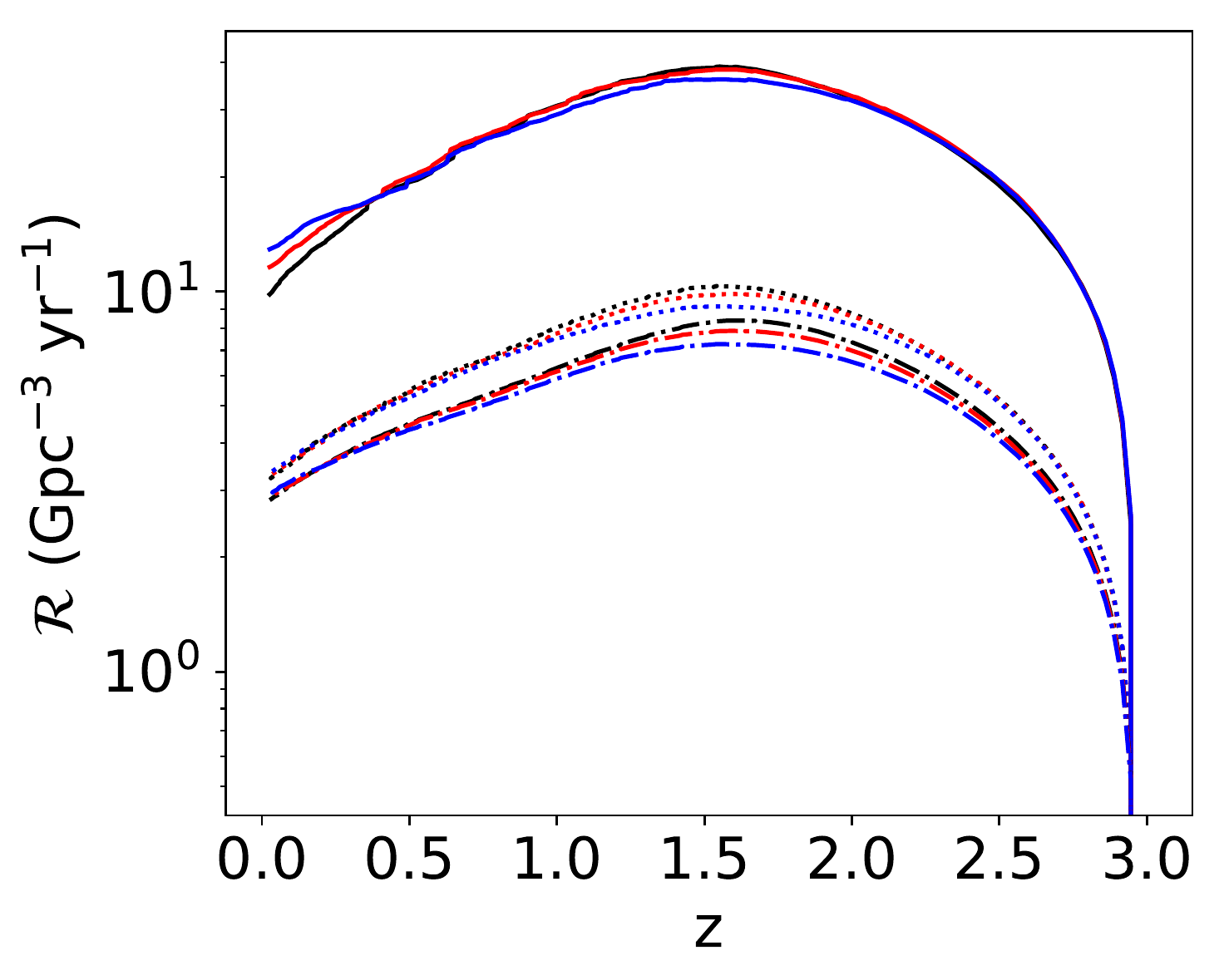}
\end{minipage}
\caption{Similar to Figure~\ref{fig:tde_singlegal}, showing the cosmological TDE rate density from MS (left) and WD (right) stars for elliptical galaxies (top) and spiral galaxies (bottom) for different Sersic indexes and mass in stars, as a function of the redshift $z$. The rates shown are reduced by the initial fraction of GCs hosting IMBHs.}
\label{fig:tde_galaver}
\end{figure*}

In the previous discussion, we assumed that there is no cusp around the IMBH. If a cusp forms and the cluster can efficiently refill it to maintain a density cusp in the IMBH vicinity, the number of stars at semi-major axis $a$ is
\begin{equation}
N(a)=N_{inf}\left(\frac{a}{R_{inf}}\right)^{3-\theta} \text{if $r<R_{inf}$}
\end{equation}
where $N_{inf}=M_{IMBH}/m_*$ is the enclosed number of stars at the influence radius and $\theta$ is the slope of the cusp. The classical result of \citet{bahcall76} predicts a slope $\theta=7/4$. The presence of a cusp may also lead to a variation in the density at the half-mass radius, which may depend on the IMBH mass via the IMBH influence radius. Heavy IMBHs can dominate the dynamical evolution of the inner cluster regions, thus possibly leading to an expansion of the cluster core and in some cases to a decrease of $\rho_h$. In the case the depletion of the cusp is more efficient than the refilling, the typical distribution of stars within the IMBH influence radius will have a smoother profile, as discussed previously. In the case of a cusp, by using Equation~(18) from \citet{syerulmer99}, we get for the TDE rate in a single cluster \citep[see also][]{stone17}
\begin{equation}
\Gamma_{\rm TDE} \approx \frac{G^{1/2} M_{GC}^{3/2}}{M_{IMBH} R_h^{3/2}}
=\frac{2\pi^{1/2}}{3^{1/2}}\frac{M_{GC}}{M_{IMBH}} \sqrt{G\rho_h} 
\ , 
\label{eqn:tde}
\end{equation}
where we used $GM_{IMBH}/R_{inf} = \sigma_c^2 \approx GM_{GC}/R_h$ and $R_h=(3M_{GC}/4\pi \rho_h)^{1/3}$. Note that the rates are higher for larger GC masses and smaller IMBH masses.

Figure \ref{fig:plum} shows the TDE rates from MS stars assuming a cusp profile (Eq. \ref{eqn:tde}) and a Plummer (Eq. \ref{eqn:tdeplu}) profile for a spiral galaxy of stellar mass $M_*=10^{11}\msun$ and Sersic index $n=2$. Near the peak of GC formation at redshift $z=3$, the TDE rate from the cusp profile exceeds the Plummer model contribution by a factor of $\sim M_{GC}/M_{IMBH}=100$. Larger TDE rates imply a more rapid consumption of the GC by the IMBH, which translates in to a lower rate at smaller redshifts. In the cusp case, the mass in stars accreted by the IMBH is larger and $M_{GC}$ will decrease faster, thus lowering the rate in Eq. \ref{eqn:tde}, which becomes comparable to the Plummer rate at lower redshifts ($z\lesssim 1.5$).

We note that in our semi-analytical approximations $\rho_h$ mainly depends on the influence of the host galaxy on the orbiting cluster, but in principle it could also depend on the details of the dynamics near the IMBH. In particular, IMBH-SBH binaries may cause an expansion of the cluster core thus decreasing to some extent $\rho_h$, which in turn may cause a more rapid cluster dissolution and reduce the expected TDE rate \citep[see e.g.,][]{bau04a,bau04b,kon13}. All these effects depend on the details of the cluster's initial properties and on the interplay between galactic tidal forces and internal cluster dynamics, which would require detailed N-body simulations.

We use the results of our simulated GC models to make predictions for the cosmological TDE rate. To compute the cosmological rate, we need information both on the cluster population as a function of redshift and host galaxy type and properties. This can be achieved by weighting the observed GC frequencies with a Schechter function, that takes into account the redshift dependence of the number density of galaxies at a given (stellar or dark matter) mass. The Schechter function has been investigated both observationally and with cosmological simulations, which seem to agree \citep{fur15}. However, observations of the cluster abundances in different galaxies is available mostly only for the local universe. \citet{rod15} used published data at $z=0$ to compute the local density of clusters \citep[see also]{har13}. To overcome this lack of data, we adopt a simple approach to compute the cosmological redshift of TDEs
\begin{equation}
\mathcal{R}(z)=n_{\mathrm{GC,total}}(z)\ \Gamma_{\mathrm{TDE}}(z)\ ,
\label{eqn:ratetde}
\end{equation}
where 
\begin{equation}\label{eqn:ngc}
n_{\mathrm{GC,total}}(z)=\kappa_1 \kappa_2(z)
\end{equation} 
is the comoving spatial density of GCs and $\Gamma_{\mathrm{TDE}}(z)$ is the average TDE rate (per cluster), computed from the results of our simulations by substituting into equation~\eqref{eqn:tdeplu}.

We define $\kappa_1$ as the local GC density that depends on the galaxy type. We correct for $\kappa_2(z)=N_{\rm GC}(z)/N_{\rm GC}(0)$, where $N_{\rm GC}(z)$ and $N_{\rm GC}(0)$ are the numbers of GCs at redshift z and GCs that survive until the present, respectively, to take into account TDEs which happen in GCs at redshift $z$ that have dissolved by $z=0$ \citep{gne14}. We then compute the contribution of each galaxy of a given mass to $\kappa_1$, by considering the relative contribution of each model weighting its present-day cluster frequency (see last column in Tab. \ref{tab:models}) by the present-day Schechter function \citep{sche76}, similar to \citet{harris16}. We adopt the Schechter function parameters as extracted from the EAGLE cosmological simulations in \citet[][see Tab.~A1]{fur15}
\begin{equation}
\Phi(M_*)=\Phi_c\left(\frac{M_*}{M_c}\right)^{-\alpha_c} e^{-M_*/M_c}\ ,
\end{equation}
where $\Phi_c$ is a normalization constant, $M_c=10^{11.14}\msun$ and $\alpha=-1.43$. In the previous equation, $M_*$ is the stellar mass of a given galaxy. We then use the stellar mass for each galaxy in our model (see Sect.~\ref{sect:galmod}) to divide the Schechter function into discrete bins ($\Delta M_{*,i}$), such that each galaxy sits in the center of its respective mass bin. Thus, the contribution of different galaxy masses to the rates are calculated by setting $\kappa_1(M_*) = \kappa_{1,av} \Phi(M_*)\Delta M_*/\int dM_* \Phi(M_*)$, where $\kappa_{1,av}=0.17$ Mpc$^{-3}$ for elliptical galaxies and $\kappa_{1,av}=0.13$ Mpc$^{-3}$ for spiral galaxies \citep{rod15}). We note that an ideal calculation of the relative contributions of different galaxies would require the coupling of the population of clusters in each galaxy with the relative galaxy numbers (given by the Schechter function) as function of redshift. Furthermore, other effects such as galaxy-galaxy mergers should also be taken into account, however, which is beyond the scope of the present paper and deserves future work.

Figure \ref{fig:tde_galaver} shows the cosmological TDE rates from MS (left) and WD (right) stars for elliptical galaxies (top) and spiral galaxies (bottom) for different Sersic indexes and masses in stars, as a function of redshift $z$. While there is no significant dependence on the Sersic index, elliptical galaxies tend to have cosmological rates $\approx 2$ times larger than spiral galaxies. In both galaxy types, the smallest galaxy has the largest rate, while the $M_*=5\times 10^{10}\msun$ and $M_*=10^{11}\msun$ galaxies give roughly similar contributions. Actually, even if the smallest galaxies have smaller numbers of clusters, they are more abundant than larger galaxies, as a consequence of the weight from the Schechter function.

We assume in all our discussions for our main model that $f=0.01$, $t_{\rm coll}=50$ Myr \citep{mil02a}, $\zeta=1$, both the IMBH and SBHs have zero spin and $e_{IMBH-SBH}=0$ (see Section~\ref{sect:evolgcimbh}). We run additional models to check how the results depend on these parameters. Importantly, among these parameters only the initial GC mass fraction in IMBHs $f$ affects the results in the cusp model (as the Plummer model is independent of the IMBH mass), since smaller IMBH masses imply larger rates (see Eq. \ref{eqn:tde}). Since we treat self-consistently the evolution of the host GC and the GC mass accreted by the IMBH, the mass in stars accreted by a less-massive IMBH could be larger and $M_{GC}$ would decrease faster at higher redshifts, thus lowering the rate (between $\propto M_{GC}$ and $M_{GC}^{2}$, see Eqs.~(\ref{eqn:rhalfm}) and Eq. \ref{eqn:tde}) at small redshifts. Apart from the very beginning, the effect of a smaller IMBH mass is compensated by a smaller GC mass, and the resulting rate is roughly independent of the IMBH mass at small redshifts. Finally, we find that only the spin may play some role at large redshifts since the GW recoil kick becomes larger and may eject more efficiently the IMBHs \citep{fragk18}, thus decreasing the rate. However, the population of IMBHs surviving within clusters is not affected since the mass ratio in a typical IMBH-SBH event is quite small and the kick velocity typically does not exceed the cluster escape speed.

We note that our results on the TDE rates correspond to the case in which all clusters in each galaxy host an IMBH in their centers. Thus these results represent an upper limit for the TDE rate from IMBHs in GCs. These numbers are consistent with the recent observations of \citet{lin18}. In reality, it may be expected that only a fraction of the clusters host IMBHs, and some form an IMBH at later times. For instance, \citet{gie15} showed that IMBH formation in star clusters is a stochastic process, with a probability of $\sim 20$\% that an IMBH will form. At the same time, IMBHs may form in the early Universe and may seed GC formation \citep{dolg17}. We found that the average TDE rate per cluster is roughly independent of the host galaxy size and properties, thus scaling by the number of clusters in a given host galaxy. If only a given fraction $\psi$ of the overall cluster population hosts an IMBH, the rate reduced in proportion to $\psi$.

\section{Gravitational Waves}
\label{sect:gw}

\begin{figure} 
\centering
\includegraphics[scale=0.58]{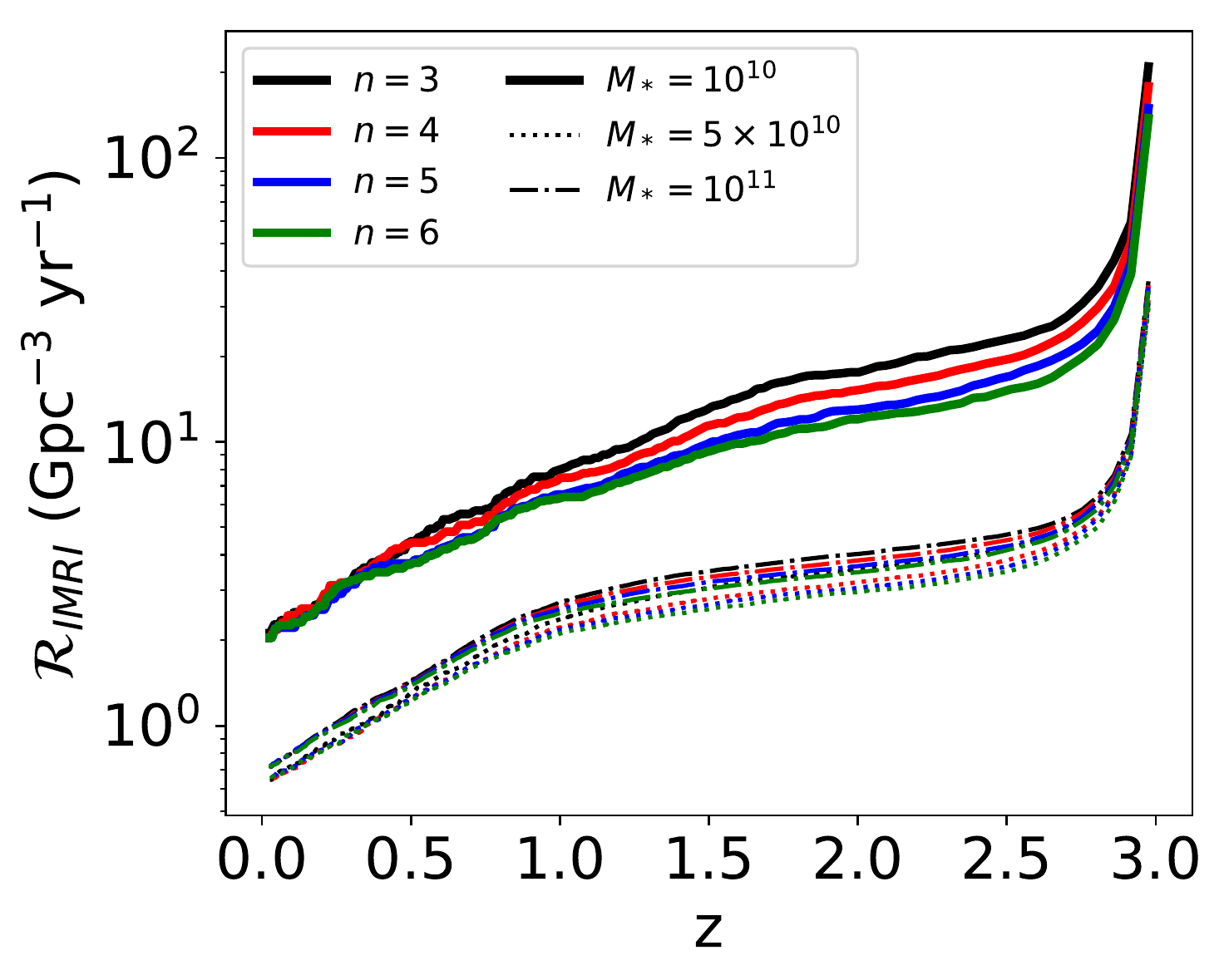}
\includegraphics[scale=0.58]{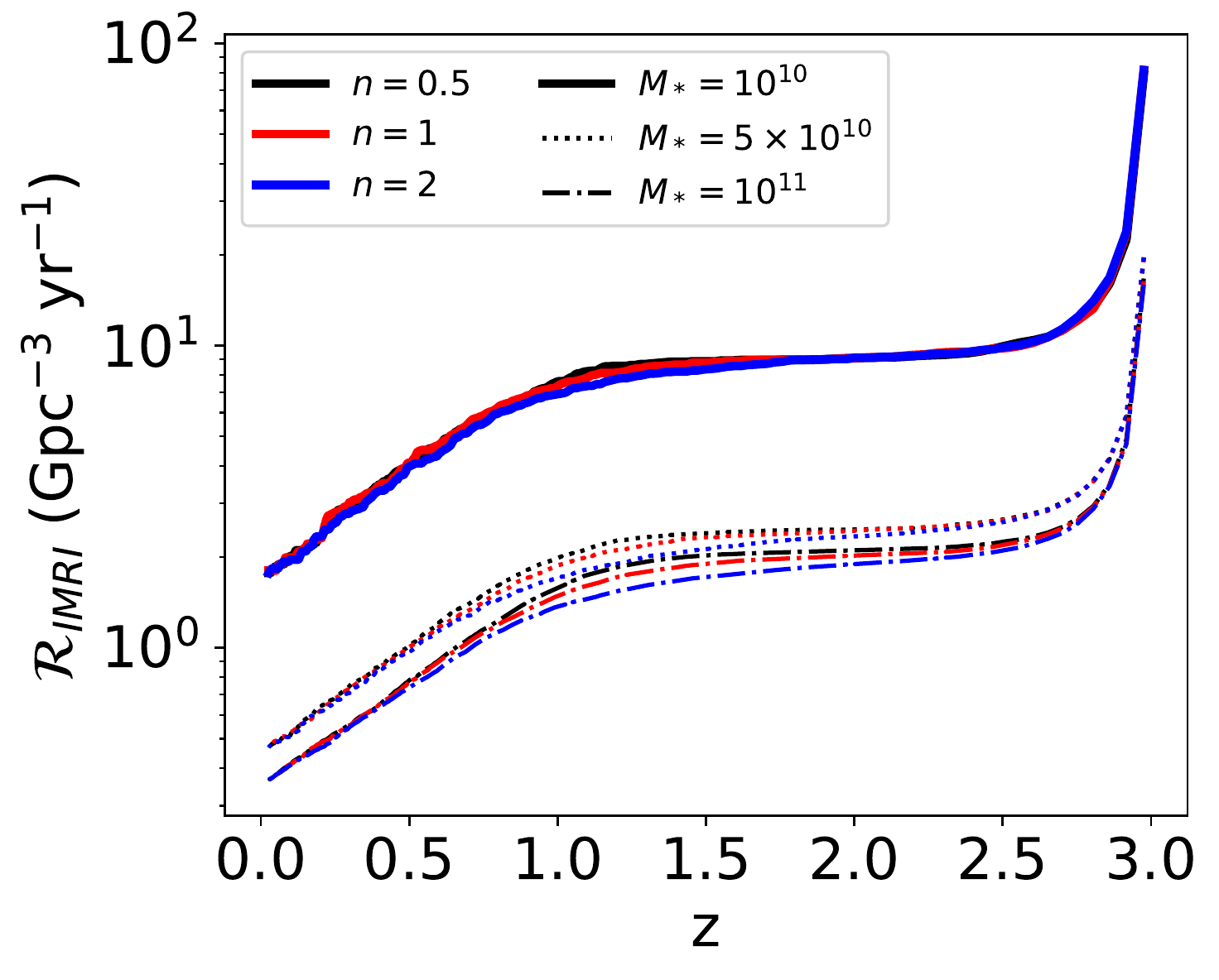}
\caption{Rate density of IMRIs from early type galaxies (top) and late type galaxies (bottom), as a function of redshift, for different galaxy stellar masses and Sersic indexes. The rates shown are reduced by the initial fraction of GCs hosting IMBHs.}
\label{fig:gw}
\end{figure}

\begin{figure} 
\centering
\includegraphics[scale=0.58]{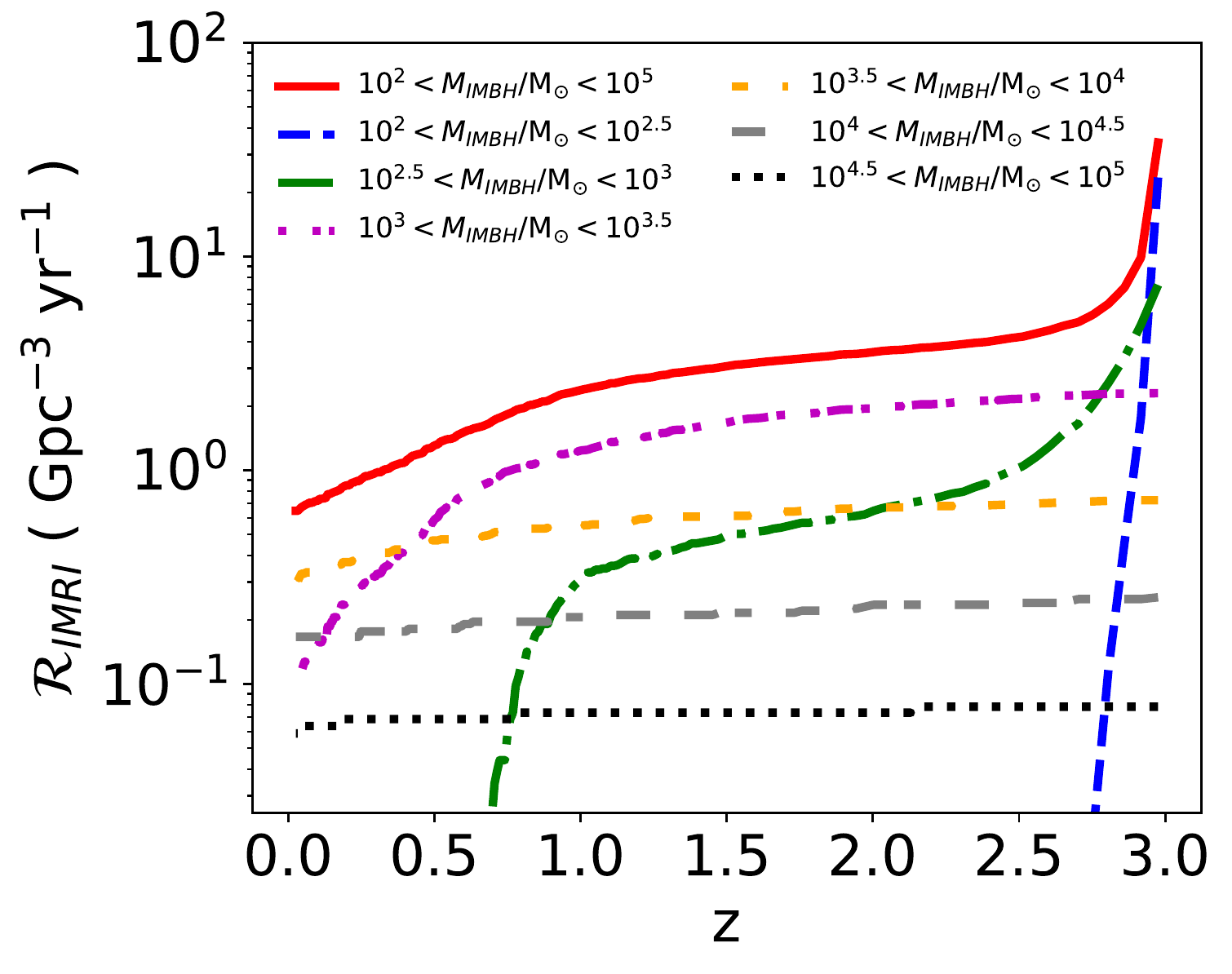}
\includegraphics[scale=0.58]{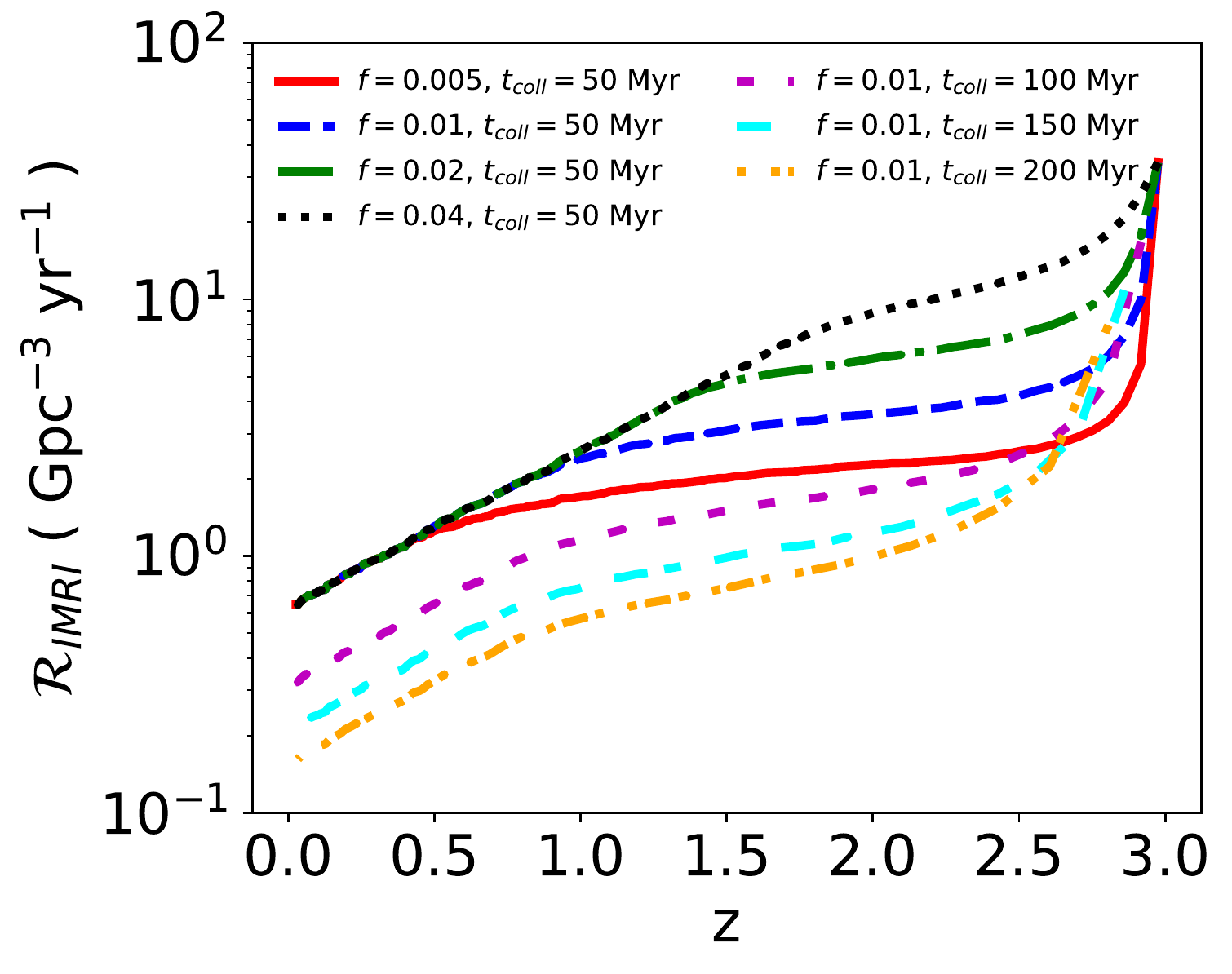}
\caption{IMRIs from a Milky-Way like galaxy. Top panel: IMRIs from IMBHs with different masses. Bottom panel: dependence on the parameters of the IMBH-SBH merger events. The rates shown are reduced by the initial fraction of GCs hosting IMBHs.}
\label{fig:gwparam}
\end{figure}

We now use the results from our simulated GC models to make predictions on the merger rate of IMRIs. As discussed in more detail in \citet{fragk18}, the IMRI rate is highly uncertain 
and depends on several parameters that describe the typical IMBH-SBH merger events. We compute the IMRI rate as
\begin{equation}
\mathcal{R_\mathrm{IMRI}}(z)=n_{\mathrm{GC,total}}(z)\ \Gamma_{\mathrm{IMRI}}(z)\ ,
\label{eqn:rategw}
\end{equation}
where $n_{\mathrm{GC,total}}(z)$ is described in Eq. \ref{eqn:ratetde} and $\Gamma_{\mathrm{IMRI}}$ is the average IMRI rate (per cluster) taken from the results of our simulations. The main parameters that affect the inferred IMRI rate are the initial fraction of GC mass in IMBHs $f=M_{\rm IMBH}/M_{\rm GC}$, the typical timescale $t_{\rm coll}$ between two subsequent IMBH-SBH mergers, the slope $\zeta$ of the SBHs mass function, the spins $\chi$ of the IMBHs and SBHs, and the eccentricities $e_{\rm IMBH-SBH}$ of the IMBH-SBH merger event. If the IMBH does not escape due to GW recoil kicks during inspiral, we generate an IMBH-SBH merger every $t_{\mathrm{coll}}$, for a total of $N_{\mathrm{coll}}=T_{\mathrm{life}}/t_{\mathrm{coll}}$, where $T_{\mathrm{life}}$ is the maximum lifetime of GCs.

Figure \ref{fig:gw} illustrates the IMRI rate from early type galaxies (top) and late type galaxies (bottom), as a function of redshift, for different galaxy stellar masses and Sersic indexes. Similarly to the TDE rate, the IMRI rate is dominated by the smallest galaxy, while no significant difference is found for different galaxy types and Sersic indexes. In our models, the IMRI rate at low redshifts is of the order of $\approx 0.5-3$ Gpc$^{-3}$ yr$^{-1}$, which becomes as high as $\approx 100$ Gpc$^{-3}$ yr$^{-1}$ near the peak of GC formation at $z=3$. Our conclusions are consistent with the recent estimates of \citet{has16} and \citet{arc18}. Yet, the exact IMRI rate remains unknown and future studies involving direct $N$-body models are needed \citep{kon13,lei14,mac16}.

Figure \ref{fig:gwparam} (top) shows the contribution to the total rate by IMBHs of different masses across cosmic time. IMBHs of mass $M_{IMBH}< 10^{3}\msun$ are efficiently ejected at large redshifts ($z\gtrsim 0.5$), while more massive IMBHs are expected to contribute to the total rate in the local universe. We divide our IMRI events in different mass bins since different instruments are expected to observe IMBH-SBH mergers with higher/lower sensitivity for different IMBH-SBH binary masses and mass ratios. The largest rate comes for $10^3\ \mathrm{M}_{\odot}<M_{\mathrm{IMBH}}\le 10^4\ \mathrm{M}_{\odot}$, which will be detectable by either \textit{LISA} or ET \citep{ama10}. ET is expected to observe GW events up to $z\approx 2$, for which our results predict a detection rate of $\approx 2$ Gpc$^{-3}$ yr$^{-1}$. Advanced LIGO, VIRGO, and KAGRA will be able to observe the low-end of the IMBH population ($\lesssim 10^3\ \mathrm{M}_{\odot}$) up to $z\approx 1.0$ \citep{abb17}. Our models predict a rate of $0.1-0.5$ Gpc$^{-3}$ yr$^{-1}$ for $0.6\lesssim z\lesssim 1$ for $300 \lesssim M_{\mathrm{IMBH}}\lesssim 1000\ \mathrm{M}_{\odot}$. \textit{LISA} may detect IMBHs of all masses, as the population of massive ($\gtrsim 10^4\ \mathrm{M}_{\odot}$) IMBHs, whose rate density is $\approx 0.1-0.2$ Gpc$^{-3}$ yr$^{-1}$, nearly independent of redshift. Finally, we note that IMBHs of a few hundred solar masses are in clusters dissolved by the galactic tidal field or are efficiently ejected by GW recoils at $z\gtrsim 2.5$. For a recent comprehensive discussion on how ground- and space-based instruments can detect GWs emitted from IMBH-SBH inspirals, see \citet{ama18}.

While in the case of TDEs the different assumptions on $f$ and $t_{\rm coll}$ do not influence significantly the relative rates (some differences arise only at high redshifts), the IMRI rate is affected by the choice of these two parameters \citep{fragk18}. Figure \ref{fig:gwparam} (bottom) shows the IMRI rate as a function of redshift for a Milky-Way like galaxy for different assumptions on $f$ and $t_{\rm coll}$ \citep{mil02a}. The former affects the rate only at large redshift since less massive IMBHs are more easily removed from GCs by GW kicks. The typical time between two mergers affects the rate at all redshifts, where larger $t_{\rm coll}$'s implies smaller IMRI rates. As noted in \citet{fragk18}, GW observations may constrain the GC models by measuring the mass, spin, and redshift distribution of IMBH mergers, and similar considerations hold for SBH-SBH mergers \citep{frak18}.

\section{Discussion and summary}
\label{sect:conclusions}

In this paper, we present a semi-analytic model that, for a given host galaxy, self-consistently models the time evolution of its globular cluster population in a realistic tidal field of the host galaxy. The model accounts for dynamical friction of the GCs, internal two-body relaxation, stellar evolution, evaporation, tidal stripping and disruption. We also take into account SBH-IMBH mergers, which can kick an IMBH out of the cluster due to a GW recoil kick. We use this model to compute the fraction of GCs still hosting IMBHs at a given redshift relative to the initial fraction, and the corresponding rate of TDE events between stars and a given IMBH. This is done for both main-sequence and white dwarf stars, such that their relative rates can be compared as a function of cosmic time as the stellar populations that comprise our model GCs age.

Our results suggest that, for a given host galaxy, the integrated TDE rate for the entire GC population can exceed that expected in the galactic nucleus due to an SMBH, for many galaxies.  This strongly argues that future time-domain surveys and observational efforts designed to identify TDE events should not confine themselves to galactic nuclei alone, but should also consider the outer galactic halo where massive old GCs hosting IMBHs are most likely to reside.  Such candidate TDE events may already have been observed, called "calcium-rich gap transients", which are associated with large physical offsets from their host galaxies \citep{frohmaier18}. The TDE recently observed by \citet{lin18} is consistent with an IMBH in an off-centre star cluster, at a distance of $\sim 12.5$ kpc from the center of the host galaxy.

Interestingly, we find that the relative rates of WD and MS TDEs change substantially over the course of cosmic time, such that the WD TDE rate can even exceed the MS TDE rate at $z \sim 0$.  Observationally, this could be interesting for a few reasons.  First, theoretical studies have shown that WD TDEs tend to be associated with strong x-ray emission \citep{metzger12,kawana18}, whereas MS TDEs are not accompanied by x-rays.  Naively, our results therefore predict that a larger fraction of TDEs associated with x-ray emission should occur at smaller redshifts in GCs, due to the increased presence of WDs as the GCs age.  This is especially interesting when convolved with the results of \citet{lei14}, who showed using a combination of $N$-body simulations and analytic methods that GCs hosting IMBHs have a low probability of simultaneously hosting x-ray binaries.  It follows that the probability of a bright x-ray source in a GC hosting an IMBH is also low.  Consequently, this contributes to a substantial increase in the probability of actually observing the x-rays associated with WD TDE events in old GCs hosting IMBHs, since other strong x-ray sources that could occult the TDE event are much less likely to be present in these GCs. Our results suggest that careful consideration of the relative rates of MS and WD TDEs in GCs across cosmic time could be worth pursuing in future studies, however we emphasize that the exact values of the relative rates are sensitive to the assumed initial stellar mass function. Also the amount of BHs in the core of the cluster may affect the results, which deserves future attention.

We also calculated the relative rate of IMBH-SBH mergers across cosmic time. We find that the typical IMRI rate at low redshift is of the order of $\approx 0.5-3$ Gpc$^{-3}$ yr$^{-1}$, which becomes as high as $\approx 100$ Gpc$^{-3}$ yr$^{-1}$ near the peak of GC formation at $z=3$. We have also shown that the largest rate comes for $10^3\ \mathrm{M}_{\odot}<M_{\mathrm{IMBH}}\le 10^4\ \mathrm{M}_{\odot}$, detectable by either \textit{LISA} or ET, while Advanced LIGO, VIRGO, and KAGRA will be able to observe the low-end of the IMBH population ($\lesssim 10^3\ \mathrm{M}_{\odot}$) up to $z\approx 1.0$, whose expected rate is $\approx 0.1-0.5$ Gpc$^{-3}$ yr$^{-1}$ for $0.6\lesssim z\lesssim 1$. Finally, we predict a rate density of $\approx 0.1-0.2$ Gpc$^{-3}$ yr$^{-1}$ for the population of massive ($\gtrsim 10^4\ \mathrm{M}_{\odot}$) IMBHs, nearly independent of redshift, observable by \textit{LISA}.

We note that in our GC models we neglected the effects of tidal shocks and eccentric orbits. As discussed, both of them would affect cluster evolution by enhancing mass-loss and shortening the typical dynamical friction timescale. Both of these effects affect mostly the clusters in the innermost regions of the host galaxy. While this effect would decrease the average TDE and IMRI rates, we argue that their effects may not be so drastic since most of the TDE and IMRI signals are generated in clusters at $\sim 2-15$ kpc from the host galaxy centre. While quantifying the effects of shocks and eccentric orbits is beyond the scope of the present paper, these topics deserve further consideration in future work.

Finally, we note that in our calculations we assume that all clusters in a given galaxy host an IMBH in their centres, making our calculations an upper limit for the TDE and GW rates from IMBHs in GCs. Probably, only a fraction of the clusters may host IMBHs, and some of them could develop an IMBH at later times \citep{gie15}. However, we find that the average TDE and IMRI rates per cluster are roughly independent of the host galaxy size and properties, thus scaling by the number of clusters in a given host galaxy. If only a given fraction $\psi$ of the overall cluster population hosts an IMBH, the rate is reduced in proportion to $\psi$.

\section{Acknowledgements}

We thank Oleg Gnedin, Re'em Sari and Tsvi Piran for useful discussions and comments. This research was partially supported by an ISF and an iCore grant. GF is supported by the Foreign Postdoctoral Fellowship Program of the Israel Academy of Sciences and Humanities. GF also acknowledges support from an Arskin postdoctoral fellowship and Lady Davis Fellowship Trust at the Hebrew University of Jerusalem. GF acknowledges the hospitality from the Center for Computational Astrophysics at Simons Foundation (New York), where the early plan of this work was conceived.  This project has received funding from the European Research Council (ERC) under the European Union's Horizon 2020 Programme for Research and Innovation ERC-2014-STG under grant agreement No. 638435 (GalNUC) and from the Hungarian National Research, Development, and Innovation Office under grant NKFIH KH-125675 (to BK). 

\bibliographystyle{yahapj}
\bibliography{refs}

\end{document}